\documentclass[12pt]{article}
\usepackage{amsmath}
\usepackage{epsfig}
\usepackage{psfrag}

\def\no{\nonumber}
\def\eps{\varepsilon}
\def\Li{\text{Li}}
\def\S{\text{S}}
\def\as{\alpha_s}
\def\Re{\text{Re}}

\def\LQCD{\Lambda_\text{QCD}}
\def\MSbar{\ensuremath{\overline{\text{MS}}}}
\def\calO{\mathcal{O}}
\def\calC{\mathcal{C}}
\def\calH{\mathcal{H}}
\def\ub{\bar{u}}
\def\ubar{\bar{u}}
\def\gev{\text{GeV}}

\allowdisplaybreaks[3]

\begin{document}


\begin{titlepage}

\begin{flushright}
 TTP09-02\\
 SFB/CPP-09-14
\end{flushright}
\vskip 2.2cm

\begin{center}
\Large\bf\boldmath
NNLO vertex corrections in charmless\\
hadronic $B$ decays: Real part \unboldmath

\normalsize
\vskip 1.5cm

{\sc Guido~Bell\footnote{E-mail:bell@particle.uni-karlsruhe.de}}

\vskip .5cm

{\it Institut f\"ur Theoretische Teilchenphysik, \\
Universit\"at Karlsruhe, D-76128 Karlsruhe, Germany}

\vskip 1.5cm

\end{center}

\begin{abstract}
\noindent
We compute the real part of the 2-loop vertex corrections for charmless
hadronic $B$ decays, completing the NNLO calculation of the topological
tree amplitudes in QCD factorization. Among the technical aspects we
show that the hard-scattering kernels are free of soft and collinear
infrared divergences at the 2-loop level, which follows after an
intricate subtraction procedure involving evanescent four quark
operators. The numerical impact of the considered corrections is found
to be moderate, whereas the factorization scale dependence of the
topological tree amplitudes is significantly reduced at NNLO. We in
particular do not find an enhancement of the phenomenologically
important ratio $|C/T|$ from the perturbative calculation.        
\end{abstract}

\vfill

\end{titlepage}

\setcounter{footnote}{0}
\renewcommand{\thefootnote}{\arabic{footnote}}


\section{Introduction}


The study of hadronic $B$ meson decays into a pair of light (charmless)
mesons reveals interesting information about the underlying four quark
interactions and the related phenomenon of CP violation. While these
decay modes are intensively investigated at current and future $B$
physics experiments, the main challenge for precise theoretical
predictions consists in the computation of the hadronic matrix
elements. QCD factorization~\cite{BBNS}, or its field theoretical
formulation in the language of Soft-Collinear Effective
Theory~\cite{SCET}, is a systematic framework to compute these matrix
elements from first principles. The starting point is a factorization
formula, which holds in the heavy quark limit $m_b\to\infty$,    
\begin{align}
\langle M_1 M_2 | Q_i | \bar{B} \rangle \;
& \simeq \; F_+^{B M_1}(0) \; f_{M_2} \int du \; \; T_{i}^I(u) \;
  \phi_{M_2}(u) \no \\
& \quad \; + \hat{f}_{B} \, f_{M_1} \, f_{M_2} \int d\omega dv du \; \;
  T_{i}^{II}(\omega,v,u) \; \phi_B(\omega) \; \phi_{M_1}(v) \;
  \phi_{M_2}(u),
\label{eq:QCDF}
\end{align}
where the perturbatively calculable hard-scattering kernels $T_i^{I,II}$
encode the short-distance strong-interaction effects and the
non-perturbative physics is confined to some process-independent
hadronic parameters such as decay constants $f_M$, light-cone
distribution amplitudes $\phi_M$ and a transition form factor $F_+^{B
  M_1}$ at maximum recoil $q^2=0$.   

In this work we address perturbative corrections to the factorization
formula (\ref{eq:QCDF}). Whereas next-to-leading order (NLO) corrections
to the hard-scattering kernels $T_i^{I,II}$ are known from the
pioneering work in~\cite{BBNS}, partial next-to-next-to-leading order
(NNLO) corrections have recently been worked
out~\cite{SpecScat:NLO:tree,SpecScat:NLO:penguin, GB:Im, GB:thesis}. The
$\as^2$ corrections to the kernels $T_i^{II}$ (\emph{spectator
  scattering}) are by now completely determined to NNLO: the corrections
for the topological tree amplitudes have been computed
in~\cite{SpecScat:NLO:tree} and the ones for the so-called penguin
amplitudes in~\cite{SpecScat:NLO:penguin}.   

In contrast to this the computation of $\as^2$ corrections to the
kernels $T_i^{I}$ (\emph{vertex corrections}) is to date
incomplete. Whereas we computed the imaginary parts of the
hard-scattering kernels for the topological tree amplitudes
in~\cite{GB:Im,GB:thesis}, we complete the NNLO calculation of the tree
amplitudes in this work by computing the respective real parts. Partial
results of this calculation, in particular the analytical expressions of
the required 2-loop Master Integrals, have already been given
in~\cite{GB:thesis}.   

The organization of this paper is as follows: The technical aspects of
the NNLO calculation are presented in Section~\ref{sec:nnlo}. We start
by briefly recalling our definitions and conventions and make some
remarks concerning the computation of the 2-loop diagrams. We then show
in some detail how to extract the hard-scattering kernels from the
matrix elements which are formally infrared divergent. This subtraction
procedure, which becomes particularly involved for the colour-suppressed
tree amplitude, is complicated due to the presence of evanescent four
quark operators which arise in intermediate steps of the
calculation. Our analytical results for the hard-scattering kernels are
summarized in Section~\ref{sec:results}. We briefly discuss the
numerical impact of the considered NNLO corrections in
Section~\ref{sec:numerics}, before we conclude in
Section~\ref{sec:conclusion}. Several technical issues of the
calculation and the explicit expressions of the hard-scattering kernels
are relegated to the Appendix.


\section{NNLO calculation}

\label{sec:nnlo}

The calculation of the real parts of the topological tree amplitudes
proceeds along the same lines as the one of the imaginary parts that we
presented in~\cite{GB:Im}. Still, the current calculation turns out to
be considerably more complex in several respects. First, it requires the
calculation of a larger amount of 2-loop integrals, which are in
addition more complicated since they involve up to three (instead of
one) massive propagators. Second, the renormalization procedure and the
infrared subtractions reveal their full 2-loop complexity only in the
current calculation as a consequence of the fact that the tree level
contribution is real. In the following we summarize the technical
aspects of the calculation and refer for a more detailed description of
the general strategy to~\cite{GB:Im} (cf.~also~\cite{GB:thesis}). 

\subsection{Operator basis}

The topological tree amplitudes can be derived from the hadronic matrix
elements of the current-current operators in the effective weak
Hamiltonian  
\begin{align}
\mathcal{H}_\text{eff} =
    \frac{G_F}{\sqrt{2}} \; V_{ud}^* V_{ub} \;
    \left( C_1 Q_1 + C_2 Q_2 \right)
    + \text{h.c.}
\end{align}
As we apply Dimensional Regularization\footnote{We write $d=4-2\eps$ and
  use an anticommuting $\gamma_5$ according to the NDR scheme.} (DR) to
regularize ultraviolet (UV) and infrared (IR) singularities, evanescent
four-quark operators appear in intermediate steps of the
calculation. The full operator basis required for the present
calculation becomes\footnote{This operator basis has been named
  \emph{CMM basis} in~\cite{GB:Im} (denoted by a hat).} 
\begin{align}
Q_1 &=
  \left[\bar u \gamma^\mu L\, T^A b\right] \;
  \left[\bar d \gamma_\mu L\, T^A u\right],\no\\
Q_2 &=
  \left[\bar u \gamma^\mu L\, b\right] \;
  \left[\bar d \gamma_\mu L\, u\right],\no\\
E_1 &=
  \left[\bar u \gamma^\mu\gamma^\nu\gamma^\rho L\, T^A b \right] \;
  \left[\bar d \gamma_\mu\gamma_\nu\gamma_\rho L\, T^A u \right]
  - 16 \,Q_1,\no\\
E_2 &=
  \left[\bar u \gamma^\mu\gamma^\nu\gamma^\rho L\, b \right] \;
  \left[\bar d \gamma_\mu\gamma_\nu\gamma_\rho L\, u \right]
  -16 \,Q_2, \no\\
E_1' &=
  \left[\bar u \gamma^\mu\gamma^\nu\gamma^\rho\gamma^\sigma\gamma^\tau
    L\, T^A b\right] \;
  \left[\bar d \gamma_\mu\gamma_\nu\gamma_\rho\gamma_\sigma\gamma_\tau
    \,L\, T^A u\right]
  -20 \,E_1 - 256 \, Q_1, \no\\
E_2' &=
  \left[\bar u \gamma^\mu\gamma^\nu\gamma^\rho\gamma^\sigma\gamma^\tau
    L\, b\right] \;
  \left[\bar d \gamma_\mu\gamma_\nu\gamma_\rho\gamma_\sigma\gamma_\tau
    \,L\, u\right]
  -20 \,E_2 - 256 \, Q_2,
\label{eq:CMMBasis}
\end{align}
with colour matrices $T^A$ and $L=1-\gamma_5$. We stress that previous
studies within QCD factorization, as
e.g.~\cite{BBNS,SpecScat:NLO:tree,SpecScat:NLO:penguin}, have often been
formulated in a different operator basis with a Fierz-symmetric
definition of the physical operators. As has been argued
in~\cite{GB:Im}, it is more convenient for the current calculation to
use the operator basis (\ref{eq:CMMBasis}) since it allows to work with
a naive anticommuting $\gamma_5$ beyond NLO~\cite{Chetyrkin:1997gb}. 

There are two different insertions of a four-quark operator which are
illustrated in Figure~1 of~\cite{GB:Im}. The first one gives rise to the
colour-allowed tree amplitude $\alpha_{1}(M_1 M_2)$, which corresponds
to the flavour content $[\bar{q}_s b]$ of the decaying $\bar B$ meson,
$[\bar{q}_s u]$ of the recoil meson $M_1$ and $[\bar{u} d]$ of the
emitted meson $M_2$. The colour-suppressed tree amplitude
$\alpha_{2}(M_1 M_2)$ follows from the second insertion and belongs to
the flavour contents $[\bar{q}_s b]$, $[\bar{q}_s d]$ and $[\bar{u} u]$,
respectively. In~\cite{GB:Im} we did not consider the second type of
insertions since we could derive the imaginary part of the
colour-suppressed amplitude from the one of the colour-allowed amplitude
using Fierz-symmetry arguments\footnote{To do so we introduced a second
  operator basis named \emph{traditional basis} in~\cite{GB:Im} (denoted
  by a tilde).}.  

In the current calculation we cannot proceed along the same lines, since
a Fierz-symmetric operator basis has not yet been worked out to
NNLO\footnote{We emphasize that the operator basis from Section~8
  in~\cite{Gorbahn:2004my} is not Fierz-symmetric and the one from
  Appendix~A in~\cite{Buras:2006gb} is presumably not
  either~\cite{Martin}.}. We therefore consider both types of insertions
in this work, which also provides an independent cross-check of our
previous result for the imaginary part of the colour-suppressed tree
amplitude.

\subsection{2-loop calculation}
\label{sec:2loop}

\begin{figure}[t!]\vspace{3mm}
\centerline{\parbox{14cm}{\centerline{
\includegraphics[width=14cm]{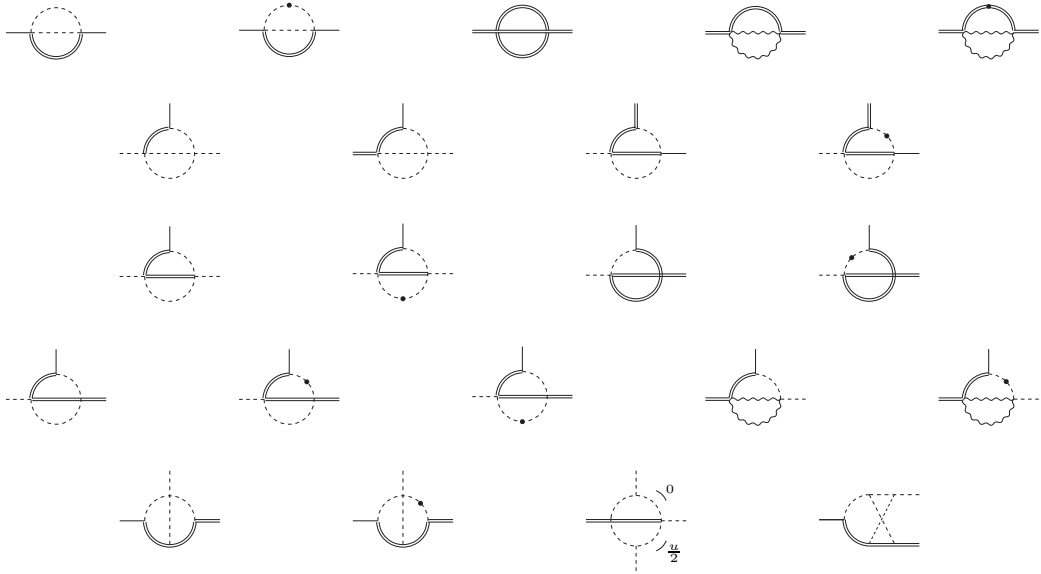}}\vspace{5mm}
\caption{\label{fig:MIs} \small \textit{Additional Master Integrals that
    appear in the calculation of the real parts of the NNLO vertex
    corrections. Dashed/double/wavy internal lines denote propagators
    with mass $0\,/\,m_b\,/\,m_c$. Dashed/solid/double external lines
    correspond to virtualities $0\,/\,u m_b^2\,/\,m_b^2$. Dotted
    propagators are taken to be squared.}}}} 
\end{figure}

The main task of the calculation consists in the computation of a large
number of 2-loop diagrams (shown in Figure~2 of~\cite{GB:Im}). We use an
automatized reduction algorithm, which is based on integration-by-parts
techniques~\cite{IBP}, to express these diagrams in terms of an
irreducible set of Master Integrals (MIs). In addition to the MIs that
appeared in the calculation of the imaginary part of the NNLO vertex
corrections (cf.~Figure~3 of~\cite{GB:Im}), we find 22 MIs which are
shown in Figure~\ref{fig:MIs}. In total the current calculation requires
the computation of 36 MIs to up to five orders in the $\eps$-expansion. 

Apart from the MIs that involve the charm quark mass, the analytical
results for the MIs from Figure~\ref{fig:MIs} can be found
in~\cite{GB:thesis}\footnote{Part of these results have recently been
  confirmed by various
  groups~\cite{Bonciani:2008wf,Beneke:2008ei}.}. The MIs can be
expressed in terms of Harmonic Polylogarithms (HPLs)~\cite{HPLs} of
weight $w\leq4$,  
\begin{align}
H(0;x) &= \ln(x),  &
H(0,0,1;x) &= \Li_{3}(x),
\no \\
H(1;x) &= -\ln(1-x), &
H(0,1,1;x) &= \S_{1,2}(x),
\no \\
H(-1;x) &= \ln(1+x), &
H(0,0,0,1;x) &= \Li_4(x),
\no \\
H(0,1;x) &= \Li_2(x), &
H(0,0,1,1;x) &= \S_{2,2}(x),
\no \\
H(0,-1;x) &= -\Li_2(-x), &
H(0,1,1,1;x) &= \S_{1,3}(x).
\no \\
H(-1,0,1;x ) &\equiv \calH_1(x),  &
H(0,-1,0,1;x ) &\equiv \calH_2(x),
\end{align}
where we introduced a shorthand notation for the last two
HPLs\footnote{The explicit expression of $\calH_1(x)$ in terms of
  Nielsen Polylogarithms can be found e.g.~in equation (10)
  of~\cite{GB:Inclusive}. On the other hand $\calH_2(x)$ has to be
  evaluated numerically (in Section~\ref{sec:convolutions} we find,
  however, analytical expressions in the convolutions with the
  light-cone distribution amplitude of the meson $M_2$).}. Moreover, the
massive non-planar 6-topology MI (last diagram from
Figure~\ref{fig:MIs}) involves a constant in the finite term which,
until recently, was only known numerically, $\calC_0
=-60.2493267(10)$~\cite{Beneke:2008ei}. In a recent work it was shown
that its analytical value is $\calC_0
=-167\pi^4/270$~\cite{Huber:2009se}. 

The charm mass dependent MIs can be found
in~\cite{GB:thesis,GB:Inclusive}. In this case there exist analytical
results apart from the finite terms of two 4-topology MIs. We may,
however, evaluate these contributions numerically to implement charm
mass effects in the current analysis.

\subsection{Renormalization}

The calculation of the renormalized matrix elements requires standard
counterterms from QCD and the effective Hamiltonian. We write the
renormalized matrix elements as  
\begin{align}
\langle Q_{i} \rangle &= Z_\psi \, Z_{i j} \,
\langle Q_{j} \rangle_\text{bare},
\end{align}
where $Z_\psi$ contains the wave-function renormalization factors of the
quark fields and $Z$ is the operator renormalization matrix in the
effective theory. Here and below we introduce a shorthand notation for
the perturbative expansions,  
\begin{align}
\langle Q_{i} \rangle_\text{(bare)} =
    \sum_{k=0}^\infty \left( \frac{\as}{4\pi} \right)^k
    \langle Q_{i} \rangle_\text{(bare)}^{(k)},
\hspace{1.5cm}
Z_{i j} = \delta_{ij} +
\sum_{k=1}^\infty  \left( \frac{\as}{4\pi} \right)^k Z_{ij}^{(k)}.
\end{align}
It turns out that the wave-function renormalization factors in $Z_\psi$
can be neglected in the calculation of the hard-scattering kernels since
they are absorbed by the form factor and the light-cone distribution
amplitude in the factorization formula, which are defined in terms of
full QCD fields (rather than HQET or SCET fields), for details
cf.~Section~4.2 of~\cite{GB:Im}. We renormalize the coupling constant in
the \MSbar-scheme, 
\begin{align}
Z_g^{(1)} = - \left( \frac{11}{6}C_A - \frac13 n_f \right)
\frac{1}{\eps},
\end{align}
and the $b$-quark mass in the on-shell scheme, 
\begin{align}
Z_{m}^{(1)} = - C_F \left( \frac{e^{\gamma_E}\mu^2}{m_b^2} \right)^\eps
\Gamma(\eps) \; \frac{3-2\eps}{1-2\eps}.
\end{align}
The 1-loop and 2-loop \MSbar~operator renormalization matrices can be
inferred from~\cite{Gorbahn:2004my,Gambino:2003zm}  
\begin{align}
Z^{(1)} &= \left(
\begin{array}{c c c c c c}
\rule[-2mm]{0mm}{7mm} -2 & \frac43 & \frac{5}{12} & \frac{2}{9} & 0 &
0\\ 
\rule[-2mm]{0mm}{7mm} 6 & 0 & 1 & 0 & 0 & 0
\end{array}
\right) \, \frac{1}{\eps},
\no\\
Z^{(2)} &= \left(
\begin{array}{c c c c c c}
\rule[-2mm]{0mm}{7mm}
17 - \frac23 n_f &
-\frac{26}{3} + \frac49 n_f &
-\frac{25}{6} + \frac{5}{36} n_f
& - \frac{31}{18} + \frac{2}{27} n_f &
\frac{19}{96} &
\frac{5}{108} \\
\rule[-2mm]{0mm}{7mm}
-39 + 2 n_f &
4 &
-\frac{31}{4} + \frac13 n_f &
0 &
\frac{5}{24} &
\frac19
\end{array}
\right) \, \frac{1}{\eps^2} \no \\
& \quad + \left(
\begin{array}{c c c c c c}
\rule[-2mm]{0mm}{7mm}
\frac{79}{12} + \frac49 n_f &
-\frac{205}{18} + \frac{10}{27} n_f &
\frac{1531}{288} - \frac{5}{216} n_f &
-\frac{1}{72} - \frac{1}{81} n_f &
\frac{1}{384} &
-\frac{35}{864}  \\
\rule[-2mm]{0mm}{7mm}
\frac{83}{4} + \frac53 n_f &
3 &
\frac{119}{16} - \frac{1}{18} n_f &
\frac89 &
-\frac{35}{192} &
-\frac{7}{72}
\end{array}
\right) \, \frac{1}{\eps},
\end{align}
where the lines refer to the physical operators and the columns to the
full operator basis including the evanescent operators from
(\ref{eq:CMMBasis}).  

\subsection{IR subtractions}
\label{sec:IR}

In order to extract the hard-scattering kernels $T_i$ we rewrite the
renormalized matrix elements in the factorized form  
\begin{align}
\langle Q_{i} \rangle &= F \cdot T_i \otimes \Phi + \ldots
\label{eq:fact:schematic}
\end{align}
where $F$ denotes the form factor, $\Phi$ the product of decay constant
and distribution amplitude, $\otimes$ the convolution integral and the
ellipsis the spectator scattering term which we disregard in the
following. As has been discussed in detail in Section~4.2
of~\cite{GB:Im}, only naively non-factorizable (nf) 1-loop diagrams
contribute to the NLO kernels,  
\begin{align}
    \langle Q_{i} \rangle_\text{nf}^{(1)}
    + Z_{ij}^{(1)} \langle Q_{j} \rangle^{(0)}
    &= F^{(0)} \cdot T_i^{(1)} \otimes \Phi^{(0)}.
\label{eq:fact:NLO}
\end{align}
Similarly, the calculation of the NNLO kernels involves only
non-factorizable 2-loop diagrams (but factorizable (f) 1-loop diagrams),
\begin{align}
    & \langle Q_{i} \rangle_\text{nf}^{(2)}
    + Z_{ij}^{(1)} \left[ \langle Q_{j}
      \rangle_\text{nf}^{(1)} + \langle Q_{j}
      \rangle_\text{f}^{(1)} \right]
    + Z_{ij}^{(2)}  \langle Q_{j} \rangle^{(0)}  \no \\
    & \qquad = F^{(0)} \cdot T_i^{(2)} \otimes \Phi^{(0)}
        + F_\text{amp}^{(1)} \cdot T_i^{(1)} \otimes \Phi^{(0)}
        + F^{(0)} \cdot T_i^{(1)} \otimes \Phi_\text{amp}^{(1)},
\label{eq:fact:NNLO}
\end{align}
where the subscript ''amp'' (amputated) has been introduced to denote
corrections without wave-function renormalization. We see that the
calculation of the NNLO kernels requires the NLO kernels to
$\calO(\eps^2)$ as they enter (\ref{eq:fact:NNLO}) in combination with
the IR-divergent form factor correction
$F_\text{amp}^{(1)}\sim1/\eps_\text{IR}^2$. As a consequence the
factorization formula has to be extended in intermediate steps of the
calculation to include evanescent operators, which have to be
renormalized such that their (IR-finite) matrix elements vanish (for
details cf.~Section 4.3 of~\cite{GB:Im}). 

At NNLO the subtraction procedure becomes somewhat involved. It is
particularly complicated in the calculation of the colour-suppressed
tree amplitude, where a Fierz-evanescent operator appears at tree
level. In the following we discuss the subtraction procedure in some
detail. Throughout this section we concentrate on the real parts of the
hard-scattering kernels, since the respective imaginary parts have
already been given in~\cite{GB:Im}. We refer to
Appendix~\ref{app:coefffuncs} for the explicit expressions of the
auxiliary coefficient functions $t_i(u)$ that we introduce below.  

\subsubsection*{Colour-allowed tree amplitude}

To NNLO we find three operators that contribute to the right hand side
of (\ref{eq:fact:schematic}). In the position space representation they
correspond to products of a local heavy-to-light current $\bar{u}(x)
\Gamma_1 b(x)$ and a non-local light-quark current $\bar{d}(y)[y,x]
\Gamma_2 u(x)$, where the usual gauge link factor $[y,x]$ is
understood. We choose the basis of Dirac structures
$\Gamma_1\otimes\Gamma_2$ as\footnote{We do not consider colour-octet
  operators since their hadronic matrix elements vanish.}  
\begin{align}
\calO &=
  \left[\gamma^\mu L\right] \otimes
  \left[\gamma_\mu L\right],\no\\
\calO_E &=
  \left[\gamma^\mu\gamma^\nu\gamma^\rho L\right] \otimes
  \left[\gamma_\mu\gamma_\nu\gamma_\rho L\right]
  -16 \,\calO, \no\\
\calO_{E'} &=
  \left[\gamma^\mu\gamma^\nu\gamma^\rho\gamma^\sigma\gamma^\tau
    L\right] \otimes
  \left[\gamma_\mu\gamma_\nu\gamma_\rho\gamma_\sigma\gamma_\tau
    \,L\right]
  -20 \,\calO_E - 256 \, \calO,
\end{align}
such that the factorized hadronic matrix element of $\calO$ gives the
standard QCD form factor and the light-cone distribution amplitude of
the emitted meson $M_2$. The operators $\calO_E$ and $\calO_{E'}$ are
evanescent.  

We first compute (\ref{eq:fact:NLO}) to $\calO(\eps^2)$ to determine the
NLO kernels. We find that the colour-singlet kernels vanish,
$T_2^{(1)}=T_{2,E}^{(1)}=T_{2,E'}^{(1)}=0$, while the colour-octet
kernels become 
\begin{align}
\Re~T_1^{(1)}(u) &=
\frac{C_F}{2N_c} \left(\frac{\mu^2}{m_b^2}\right)^\eps
\bigg\{ t_0(u) -6L
+ \Big( t_1(u) + 3 L^2 \Big) \eps
+ \Big( t_2(u) - L^3 \Big) \eps^2 + \calO(\eps^3)
\bigg\},
\no \\
\Re~T_{1,E}^{(1)}(u) &=
-\frac{C_F}{4N_c} \left(\frac{\mu^2}{m_b^2}\right)^\eps
\bigg\{ t_{E,0}(u) +2L
+ \Big( t_{E,1}(u) - L^2 \Big) \eps + \calO(\eps^2)
\bigg\},
\label{eq:T1:a1}
\end{align}
and $T_{1,E'}^{(1)}=0$ with $L=\ln \mu^2/m_b^2$. The IR subtractions on
the right hand side of (\ref{eq:fact:NNLO}) require in addition form
factor and wave function corrections to the operators $\calO$ and
$\calO_E$ (they can be found in Section 4.3 of~\cite{GB:Im}). We finally
perform the convolutions of the NLO kernels with the wave function
corrections, which yields  
\begin{align}
F^{(0)} \; \Re \;T_1^{(1)} \; \Phi_\text{amp}^{(1)} &=
\frac{C_F^2}{N_c} \left(\frac{\mu^2}{m_b^2}\right)^\eps
\bigg\{  \frac{t_3(u)}{\eps} + t_4(u) +\calO(\eps) \bigg\} \;
F^{(0)} \;\Phi^{(0)}
\label{eq:ConT1:a1}
\end{align}
and an additional $\mu$-dependent contribution to the physical kernel
from 
\begin{align}
F_E^{(0)} \; \Re \;T_{1,E}^{(1)} \; \Phi_{\text{amp},E}^{(1)} \;&\to\;
\frac{C_F^2}{N_c} \bigg\{ 12L+ t_{E,2}(u) +\calO(\eps) \bigg\} \;
F^{(0)} \;\Phi^{(0)}.
\label{eq:ConT1E:a1}
\end{align}

\subsubsection*{Colour-suppressed tree amplitude}

In this case we find an analogous set of operators,
\begin{align}
\tilde{\calO} &=
  \left[\gamma^\mu L\right] \tilde{\otimes }
  \left[\gamma_\mu L\right],\no\\
\tilde{\calO}_E &=
  \left[\gamma^\mu\gamma^\nu\gamma^\rho L\right] \tilde{\otimes }
  \left[\gamma_\mu\gamma_\nu\gamma_\rho L\right]
  -16 \,\tilde{\calO}, \no\\
\tilde{\calO}_{E'} &=
  \left[\gamma^\mu\gamma^\nu\gamma^\rho\gamma^\sigma\gamma^\tau
    L\right] \tilde{\otimes}
  \left[\gamma_\mu\gamma_\nu\gamma_\rho\gamma_\sigma\gamma_\tau
    \,L\right]
  -20 \,\tilde{\calO}_E - 256 \, \tilde{\calO},
\end{align}
but the fields are now given in the wrong ordering $\bar{u}(y)[y,x]
\Gamma_1 b(x)$ and $\bar{d}(x) \Gamma_2 u(x)$ (indicated by
$\tilde{\otimes}$), which does not yield a form factor and a light-cone
distribution amplitude. The latter follow from the factorized hadronic
matrix element of the operator  
\begin{align}
  \calO &= \bar{d}(x) \gamma^\mu L b(x) \;\otimes\;
  \bar{u}(y)[y,x] \gamma_\mu L u(x),
\end{align}
which is the Fierz-symmetric counterpart of $\tilde{\calO}$. We
therefore extend the right hand side of (\ref{eq:fact:schematic}) to
include four operators in this case: the physical operator $\calO$, the
evanescent operators $\tilde{\calO}_E$ and $\tilde{\calO}_{E'}$ and the
Fierz-evanescent operator $\tilde{\calO}_F\equiv\tilde{\calO} - \calO$. 

The IR subtractions turn out to be particularly complicated in this
case, due to the fact that the evanescent operator $\tilde{\calO}_F$
already appears in the tree level calculation. As a consequence the
naive split-up into non-factorizable diagrams, which contribute to the
hard-scattering kernels, and factorizable diagrams, which give form
factor and wave function corrections, is spoiled. In NLO we find that
equations (30) and (31) of~\cite{GB:Im} should be replaced
by\footnote{We introduce the ''hat'' notation to distinguish these
  quantities from those of the preceding section.}   
\begin{align}
\langle \hat{Q}_{i} \rangle_\text{nf}^{(1)}
+ Z_{ij}^{(1)} \langle \hat{Q}_{j}
\rangle^{(0)} &=
\hat{F}^{(0)} \cdot \hat{T}_i^{(1)} \otimes \hat{\Phi}^{(0)}
+ \hat{\Delta}_{F,i}^{(1)},
\no\\
\langle \hat{Q}_{i} \rangle_\text{f}^{(1)}
+ Z_\psi^{(1)} \langle \hat{Q}_{i}
\rangle^{(0)} &=
\hat{F}^{(1)} \cdot \hat{T}_i^{(0)} \otimes \hat{\Phi}^{(0)}
+ \hat{F}^{(0)} \cdot \hat{T}_i^{(0)} \otimes \hat{\Phi}^{(1)}
- \hat{\Delta}_{F,i}^{(1)},
\label{eq:fact:NLO:a2}
\end{align}
where $\hat{\Delta}_{F,i}^{(1)}$ contains the (non-vanishing) 1-loop
counterterms of the form factor and wave function corrections for the
Fierz-evanescent operator $\tilde{\calO}_F$. In other words, the
split-up in the above example of the colour-allowed tree amplitude
followed from the fact that the corresponding counter\-terms vanish for
the physical operator $\calO$ (i.e.~$\Delta_{i}^{(k)}=0$). 

From the first equation in (\ref{eq:fact:NLO:a2}) we see that we can
neglect the factorizable 1-loop diagrams in the computation of the NLO
kernels. In order to account for the counterterm contribution
$\hat{\Delta}_{F,i}^{(1)}$, we compute the UV-divergences of the 1-loop
diagrams from Figure~5 and~6 of~\cite{GB:Im} with an insertion of the
Fierz-evanescent operator $\tilde{\calO}_F$. We find that the
counterterms (ct) are given by 
\begin{align}
\hat{F}_{F}^{(1)} \,\hat{\Phi}_F^{(0)} \big|_\text{ct} (u)&=
C_F \left\{ \hat{F}^{(0)}  \,\hat{\Phi}^{(0)} (u)
+\frac{1}{4\eps}  \hat{F}_E^{(0)}  \,\hat{\Phi}_E ^{(0)} (u)\right\}, 
\no\\
\hat{F}_F^{(0)} \,\hat{\Phi}_{F}^{(1)} \big|_\text{ct} (u)&=
2C_F \int_0^1 dw \; V_E(u,w) \; \left\{ \hat{F}^{(0)}
  \,\hat{\Phi}^{(0)}(w) +\frac{1}{4\eps}  \hat{F}_E^{(0)}
  \,\hat{\Phi}_E ^{(0)} (w)\right\}, 
\label{eq:ct:1loop}
\end{align}
with $V_E(u,w)$ from equation (45) of~\cite{GB:Im}. Convoluting these
expressions with the LO kernels, $\hat{T}_{1,F}^{(0)}=C_F/N_c$ and
$\hat{T}_{2,F}^{(0)}=1/N_c$, yields the additional counterterm
contributions  
\begin{align}
\hat{\Delta}_{F,1}^{(1)} &=
C_F \hat{\Delta}_{F,2}^{(1)} =
 \frac{2 C_F^2}{N_c} \left\{ \hat{F}^{(0)}  \,\hat{\Phi}^{(0)}
+\frac{1}{4\eps}  \hat{F}_E^{(0)}  \,\hat{\Phi}_E ^{(0)} \right\}.
\label{eq:Delta1}
\end{align}
With this prescription the NLO kernels turn out to be free of
IR-singularities. Evaluating the first equation of
(\ref{eq:fact:NLO:a2}) to $\calO(\eps^2)$ gives (in terms of $T_1^{(1)}$
from (\ref{eq:T1:a1})), 
\begin{align}
\hat{T}_1^{(1)}(u) + C_F &=
-\frac{\hat{T}_{2}^{(1)}(u)}{2N_c}
\no \\
&= - \frac{T_1^{(1)}(u)}{N_c}
-\frac{C_F}{2N_c^2} \left(\frac{\mu^2}{m_b^2}\right)^\eps
\bigg\{ \Big( \hat{t}_{1}(u) +2L  \Big) \eps
+ \Big( \hat{t}_{2}(u) - L^2 \Big) \eps^2  + \calO(\eps^3)
\bigg\},
\no \\
\hat{T}_{1,E}^{(1)}(u) &=
-\frac{\hat{T}_{2,E}^{(1)}(u)}{2N_c} =
\frac{C_F}{8N_c^2} \left(\frac{\mu^2}{m_b^2}\right)^\eps
\bigg\{ 2L + \hat{t}_{E,0}(u)
+ \Big( \hat{t}_{E,1}(u)  - L^2 \Big) \eps + \calO(\eps^2)
\bigg\},
\no \\
\hat{T}_{1,F}^{(1)}(u) &=
-\frac{\hat{T}_{2,F}^{(1)}(u)}{2N_c} =
- \frac{T_1^{(1)}(u)}{N_c}
-\frac{C_F}{2N_c^2} \left(\frac{\mu^2}{m_b^2}\right)^\eps
\bigg\{2 + \hat{t}_1(u) \eps + \calO(\eps^2) \bigg\}.
\label{eq:T12:a2}
\end{align}
We next compute form factor and wave function corrections to $\calO$,
$\tilde{\calO}_E$ and $\tilde{\calO}_F$. Proceeding along the lines
outlined in Section 4.3~of~\cite{GB:Im}, we obtain\footnote{The
  corrections to the physical operator $\calO$ can be found
  in~\cite{GB:Im}.}   
\begin{align}
\hat{F}_\text{amp,E}^{(1)} \,\hat{\Phi}_E^{(0)}&= C_F
\left[24- \left( \frac{e^{\gamma_E}\mu^2}{m_b^2} \right)^\eps
\Gamma(\eps)\frac{24\eps(1+\eps)}{(1-\eps)^2} \right]
\;  \hat{F}^{(0)} \,\hat{\Phi}^{(0)}
\no\\
&\quad
-C_F \left[ \frac{3}{\eps} +\left( \frac{e^{\gamma_E}\mu^2}{m_b^2}
  \right)^\eps \Gamma(\eps)
  \frac{1-6\eps+16\eps^2-14\eps^3}{\eps(1-2\eps)(1-\eps)^2} \right]
\; \hat{F}_E^{(0)} \,\hat{\Phi}_E^{(0)}
\no\\
&\quad
+C_F \left[ \frac{1}{4\eps} - \left( \frac{e^{\gamma_E}\mu^2}{m_b^2}
  \right)^\eps  \frac{\Gamma(\eps)}{4(1-\eps)^2} \right]
\; \hat{F}_{E'}^{(0)} \,\hat{\Phi}_{E'}^{(0)}
\no\\
&\quad
-C_F \left( \frac{e^{\gamma_E}\mu^2}{m_b^2} \right)^\eps \Gamma(\eps) 
\frac{24\eps(1+\eps)}{(1-\eps)^2} \;  \hat{F}_F^{(0)}
\,\hat{\Phi}_F^{(0)}, 
\no
\\
\hat{F}_\text{amp,F}^{(1)} \,\hat{\Phi}_F^{(0)}&=  C_F \left[1 - \left( 
    \frac{e^{\gamma_E}\mu^2}{m_b^2} \right)^\eps \Gamma(\eps)
  \left(\frac{1-3\eps+6\eps^2-6\eps^3}{\eps(1-2\eps)(1-\eps)^2}  -
    \frac{1-\eps+2\eps^2}{\eps(1-2\eps)} \right)
\right]\; \hat{F}^{(0)} \,\hat{\Phi}^{(0)}
\no\\
&\quad
+ C_F \left[ \frac{1}{4\eps} - \left( \frac{e^{\gamma_E}\mu^2}{m_b^2} 
  \right)^\eps 
 \frac{\Gamma(\eps)}{4(1-\eps)^2} \right] \; \hat{F}_{E}^{(0)}
\,\hat{\Phi}_{E}^{(0)}
\no\\
&\quad
- C_F \left( \frac{e^{\gamma_E}\mu^2}{m_b^2} \right)^\eps\Gamma(\eps)
  \frac{1-3\eps+6\eps^2-6\eps^3}{\eps(1-2\eps)(1-\eps)^2}
\; \hat{F}_F^{(0)} \,\hat{\Phi}_F^{(0)}
\end{align}
and for the wave function corrections
\begin{align}
\hat{F}_E^{(0)} \, \hat{\Phi}_\text{amp,E}^{(1)} &=
48 C_F \bigg[ V_E \otimes \hat{F}^{(0)} \, \hat{\Phi}^{(0)} \bigg]
- \frac{2C_F}{\eps} \bigg[ \Big( V + 3 V_E \Big) \otimes \hat{F}_E^{(0)}
\, \hat{\Phi}_E^{(0)} \bigg]
\no\\
&\quad
+ \frac{C_F}{2\eps}
\bigg[ V_E \otimes \hat{F}_{E'}^{(0)} \, \hat{\Phi}_{E'}^{(0)}\bigg],
\no\\
\hat{F}_F^{(0)} \, \hat{\Phi}_\text{amp,F}^{(1)} &=
2C_F \bigg[ V_E \otimes \hat{F}^{(0)} \, \hat{\Phi}^{(0)} \bigg]
+ \frac{C_F}{2\eps} \bigg[ V_E \otimes \hat{F}_E^{(0)} \,
\hat{\Phi}_E^{(0)} \bigg]
- \frac{2 C_F}{\eps} \bigg[ V  \otimes \hat{F}_F^{(0)} \,
\hat{\Phi}_F^{(0)} \bigg], 
\end{align}
where $\otimes$ represents a convolution and $V$ is the
Efremov-Radyushkin-Brodsky-Lepage (ERBL) kernel~\cite{ERBL} (given
explicitly in equation (43) of~\cite{GB:Im}). We finally compute the
convolutions of the NLO kernels with the wave function corrections. For
the physical operator $\calO$ we get  
\begin{align}
\hat{F}^{(0)} \;\hat{T}_1^{(1)} \; \hat{\Phi}_\text{amp}^{(1)} &=
-\frac{1}{2N_c} \hat{F}^{(0)} \;\hat{T}_2^{(1)} \;
\hat{\Phi}_\text{amp}^{(1)}
\no\\
&
= -\frac{1}{N_c} F^{(0)} \;T_1^{(1)} \; \Phi_\text{amp}^{(1)}
+ \frac{C_F^2}{2N_c^2} \bigg\{ \hat{t}_3(u) + \calO(\eps) \bigg\} \;
\hat{F}^{(0)} \;\hat{\Phi}^{(0)},
\label{eq:ConT12:a2}
\end{align}
while the evanescent operators give again $\mu$-dependent corrections to
the physical kernels
\begin{align}
\hat{F}_E^{(0)} \;\hat{T}_{1,E}^{(1)} \; \hat{\Phi}_{\text{amp},E}^{(1)}
&=
- \frac{1}{2N_c} \hat{F}_E^{(0)} \;\hat{T}_{2,E}^{(1)} \;
\hat{\Phi}_{\text{amp},E}^{(1)}
\;\to\;
\frac{C_F^2}{N_c^2} \bigg\{ 6L+ \hat{t}_{E,2}(u) +\calO(\eps)
\bigg\} \;  \hat{F}^{(0)} \;\hat{\Phi}^{(0)},
\no\\
\hat{F}_F^{(0)} \;\hat{T}_{1,F}^{(1)} \; \hat{\Phi}_{\text{amp},F}^{(1)}
&=
- \frac{1}{2N_c} \hat{F}_F^{(0)} \;\hat{T}_{2,F}^{(1)} \;
\hat{\Phi}_{\text{amp},F}^{(1)}
\;\to\;
\frac{C_F^2}{2N_c^2} \bigg\{ 6L+ \hat{t}_{F,0}(u) +\calO(\eps) \bigg\} \;
\hat{F}^{(0)} \;\hat{\Phi}^{(0)}.
\label{eq:ConT12EF:a2}
\end{align}
According to (\ref{eq:fact:NNLO}) we now have assembled all pieces to
perform the IR subtractions in NNLO. However, as we have seen above in
the calculation of the NLO kernels, the naive split-up into factorizable
and non-factorizable contributions is spoiled for the colour-suppressed
amplitude. In analogy to (\ref{eq:fact:NLO:a2}) we therefore have to
account for an additional contribution $\hat{\Delta}_{F,i}^{(2)}$ on the
right hand side of (\ref{eq:fact:NNLO}), which represents the 2-loop
counterterms of the form factor and wave function corrections for the
Fierz-evanescent operator $\tilde{\calO}_F$.  

The calculation of this counterterm contribution requires a rather
complicated 2-loop calculation on its own. We refer to
Appendix~\ref{app:Delta2} for the details of this calculation and quote
the contribution to the physical kernel only,  
\begin{align}
\hat{\Delta}_{F,1}^{(2)} &=
C_F \hat{\Delta}_{F,2}^{(2)}
\;\to\;
- \frac{2 C_F^2}{N_c} \bigg\{
\frac{C_F}{\eps^2}  + \bigg[ \Big(1 + L\Big) C_F + \frac{11}{6} C_A
- \frac13 n_f \bigg] \frac{1}{\eps}
\no\\
&\hspace{2.1cm}
+ \bigg( 28 + \frac{\pi^2}{12} + 7 L + \frac12 L^2 \bigg) C_F
- \frac{149}{36} C_A - \frac{5}{18} n_f
+\calO(\eps)\bigg\}
\hat{F}^{(0)}  \,\hat{\Phi}^{(0)}.
\end{align}


\section{Vertex corrections in NNLO}

\label{sec:results}

As we have seen in the last section, the NNLO calculation of the
hard-scattering kernels requires a rather complex subtraction procedure
of UV- and IR-divergences. The fact that the kernels turn out to be free
of any singularities represents both a non-trivial confirmation of the
factorization framework and a stringent cross-check of our calculation.

\subsection{Hard-scattering kernels}

In terms of the Wilson coefficients $C_i$ of the physical operators
$Q_i$ from the operator basis (\ref{eq:CMMBasis}), the topological tree
amplitudes take to NNLO the form  
\begin{align}
\alpha_1(M_1 M_2) \;&=\;
C_2 + \frac{\as}{4\pi} \, \frac{C_F}{2 N_c}
\bigg\{ C_1 V^{(1)} + \frac{\as}{4\pi} \left[
C_{1} \, V_1^{(2)} + C_{2} \, V_2^{(2)} \right]
+ \calO(\as^2)  \bigg\} + \ldots
\no\\
\alpha_2(M_1 M_2) \;&=\;
\frac{C_F}{N_c} C_1 + \frac{C_2}{N_c}
+ \frac{\as}{4\pi} \, \frac{C_F}{2 N_c}
\bigg\{ \left( 2 C_2 - \frac{C_1}{N_c} \right) V^{(1)} -2 C_A \, C_1
\no\\
& \quad
+ \frac{\as}{4\pi} \left[
\left( 2 C_2 - \frac{C_1}{N_c} \right) V_1^{(2)}
+ \left( \frac{C_F}{N_c} C_1 + \frac{C_2}{N_c} \right) V_2^{(2)}
+ 2 C_A \, C_2 \, V^{(1)} \right.
\no\\
& \hspace{1.5cm} \left.
+\left( 8C_F-\frac{113}{18} C_A - \frac59 n_f \right) C_A \, C_1 \right]
+ \calO(\as^2)  \bigg\} + \ldots
\end{align}
where the ellipsis refer to the terms from spectator scattering which we
disregard in the following. In this notation the $\as$ corrections have
been expressed in terms of the convolution   
\begin{align}
V^{(1)}   &=
    \int_0^1 du \; \Big( -6L + g_2(u) +i \pi g_1(u)  \Big) \;
    \phi_{M_2}(u), 
\end{align}
where $L=\ln \mu^2/m_b^2$ and (recall that $\ubar=1-u$)
\begin{align}
g_1(u) &= -3 -2 \ln u + 2 \ln \ubar,
\no\\
g_2(u) &= -22 + \frac{3(1-2u)}{\ubar} \ln u + \bigg[2\Li_2(u) - \ln^2 u
- \frac{1-3u}{\ubar} \ln u -(u\to\ubar) \bigg]. 
\end{align}
If we transform these expressions into the Fierz-symmetric operator
basis that has been used in many previous QCD factorization analyses, we
reproduce the NLO result from~\cite{BBNS}. In NNLO we find the
convolutions  
\begin{align}
V_1^{(2)}  &=
    \int_0^1 du \; \bigg\{ \Big( 36 C_F -29 C_A +2 n_f \Big) L^2
\no \\
&\qquad
+  \bigg\{ \Big( \frac{29}{3} C_A - \frac23 n_f \Big) g_2(u)
-\frac{91}{6} C_A - \frac{10}{3} n_f + C_F h_6(u)
\no\\
&\hspace{1.5cm}
+ i \pi \Big[ \Big( \frac{29}{3} C_A - \frac23 n_f \Big)
g_1(u) + C_F h_1(u) \Big] \bigg\} L
\no\\
&\qquad
+ C_F h_7(u) + C_A h_8(u) + (n_f-2) h_9(u;0) + h_9(u;z) + h_9(u;1)
\no\\
&\qquad
+ i \pi \Big[
C_F h_2(u) + C_A h_3(u) + (n_f-2) h_4(u;0) + h_4(u;z) + h_4(u;1)
\Big] \bigg\}   \phi_{M_2}(u), \no \\
V_2^{(2)} &=
    \int_0^1 du \; \bigg\{ 18 L^2
+ \Big(21-6g_2(u)-6i\pi g_1(u)\Big) L + h_5(u) + i \pi h_0(u) \bigg\}
    \phi_{M_2}(u),
\label{eq:V2}
\end{align}
where $n_f=5$ represents the number of active quark flavours and
$z=m_c/m_b$. The explicit expressions for the NNLO kernels $h_{0-4}$,
which  specify the imaginary parts of the topological tree amplitudes,
can be found in~\cite{GB:Im}. As a new result we obtained the real parts
of the topological tree amplitudes to NNLO, which have been given in
terms of a new set of kernels $h_{5-9}$ that are listed in
Appendix~\ref{app:kernels}.  

Partial structures of our NNLO result can be cross-checked. First, we
verified that the scale dependence between the Wilson coefficients, the
coupling constant, the hard-scattering kernels and the light-cone
distribution amplitude cancels in the tree amplitudes $\alpha_i(M_1M_2)$
to $\calO(\as^2)$ as it should\footnote{We emphasize that this
  cancellation would have been incomplete, if $\mu$-dependent
  contributions from the mixing of evanescent operators as e.g.~in
  (\ref{eq:ConT1E:a1}) or (\ref{eq:ConT12EF:a2}) had been
  missed.}. Second, we compared the terms proportional to~$n_f$ with the
analysis of the large $\beta_0$-limit in~\cite{betazero} and found
agreement. Finally, we reproduced the imaginary part of the
colour-suppressed amplitude from our earlier analysis in~\cite{GB:Im},
which was derived on the basis of Fierz-symmetry arguments.  

\subsection{Convolutions in Gegenbauer expansion} 
\label{sec:convolutions}

We expand the light-cone distribution amplitude of the emitted meson
$M_2$ into the eigenfunctions of the 1-loop evolution kernel,  
\begin{align}
\phi_{M_2}(u) &=
     6 u \ubar \left[ 1 + \sum_{n=1}^\infty \, a_n^{M_2} \;
     C_n^{(3/2)}(2u-1) \right],
\end{align}
where $a_n^{M_2}$ and $C_n^{(3/2)}$ are the Gegenbauer moments and
polynomials, respectively. It is convenient to truncate this expansion
at $n=2$, which allows us to perform the convolution integrals in our
final expression (\ref{eq:V2}) explicitly. The convolution with the NLO
kernel results in\footnote{We refer to~\cite{GB:Im} for the convolutions
  with the kernels $g_1$ and $h_{0-4}$.}, 
\begin{align}
\int_0^1 du \; g_2(u)  \; \phi_{M_2}(u) \;&=\;
-\frac{45}{2} + \frac{11}{2} \, a_1^{M_2} - \frac{21}{20} \, a_2^{M_2},
\end{align}
whereas the convolutions with the NNLO kernels become
\begin{align}
\int_0^1 du \; h_5(u)  \; \phi_{M_2}(u) \;&=\;
 \frac{5347}{60} - \frac{14833}{5} \zeta_3 + 3744 \,\zeta_5
\no \\
& \hspace{2cm}
+ \Big( \frac{12487}{12} - 936 \,\zeta_3 + 72 \ln 2 \Big) \pi^2
+ \frac{239\pi^4}{90}
\no \\
& \quad
+ \bigg\{ \frac{4568}{15} + \frac{77157}{5} \zeta_3 - 19008 \,\zeta_5 
\no \\
&\hspace{2cm}
- \Big( \frac{21807}{4} - 4752 \,\zeta_3 + 24 \ln 2 \Big) \pi^2
- \frac{181\pi^4}{10}
 \bigg\}
\, a_1^{M_2}
\no \\
& \quad
+ \bigg\{ \frac{32369221}{12600} - \frac{2236872}{35} \zeta_3 + 74304
\,\zeta_5
\no\\
&\hspace{2cm}
+ \Big( \frac{204218}{9} - 18576 \,\zeta_3 - 2064 \ln 2\Big) \pi^2
+ \frac{797\pi^4}{10}
 \bigg\}
\, a_2^{M_2},
\no \\
\int_0^1 du \; h_6(u)  \; \phi_{M_2}(u) \;&=\;
    348 -\frac{154}{3} \, a_1^{M_2} + \frac{329}{40}\, a_2^{M_2},
\no \\
\int_0^1 du \; h_7(u)  \; \phi_{M_2}(u) \;&=\
\frac{12809}{60} - \frac{26606}{5} \zeta_3 + 6564 \, \zeta_5
\no \\
& \hspace{2cm}
+ \Big( \frac{12811}{6} - 1764 \, \zeta_3 - 48 \ln 2 \Big) \pi^2
+ \frac{134\pi^4}{45}
\no \\
& \quad
+ \bigg\{ \frac{50387}{180} + \frac{132294}{5} \zeta_3 - 32472 \,
\zeta_5
\no \\
& \hspace{2cm}
- \Big( \frac{66425}{6} - 8856 \, \zeta_3 - 1296 \ln 2 \Big) \pi^2
- \frac{176\pi^4}{5} \bigg\}
\, a_1^{M_2}
\no \\
& \quad
+ \bigg\{ \frac{75807647}{12600} - \frac{3960924}{35} \zeta_3 + 129204 
\zeta_5
\no \\
& \hspace{2cm}
+ \Big( \frac{2074841}{45} - 34884 \, \zeta_3 - 8672 \ln 2\Big)
\pi^2 + \frac{727\pi^4}{5} \bigg\}
\, a_2^{M_2},
\no \\
\int_0^1 du \; h_8(u)  \; \phi_{M_2}(u) \;&=\
- \frac{74611}{180} + \frac{618}{5} \zeta_3  - 186 \, \zeta_5
\no \\
& \hspace{2cm}
- \Big( \frac{815}{6} - 108 \, \zeta_3 - 36 \ln 2\Big) \pi^2
- \frac{169\pi^4}{120}
\no \\
& \quad
+ \bigg\{ \frac{355693}{360} + \frac{10818}{5} \zeta_3 - 2556 \, \zeta_5
\no \\
& \hspace{2cm}
- \Big( \frac{1081}{12} - 270 \, \zeta_3 + 684 \ln 2\Big) \pi^2
+ \frac{151\pi^4}{8} \bigg\}
\, a_1^{M_2}
\no \\
& \quad
+ \bigg\{ -\frac{148920211}{25200} + \frac{128283}{35} \zeta_3 - 666
\zeta_5
\no \\
& \hspace{2cm}
- \Big( \frac{66545}{18} - 1458 \, \zeta_3 - 4120 \ln 2\Big)
\pi^2 - \frac{1403\pi^4}{20} \bigg\}
\, a_2^{M_2}.
\end{align}
We finally perform the convolution with the hard-scattering kernel
$h_9(u;z_f)$, which stems from the diagrams with a closed fermion
loop. As this contribution depends on the mass $m_f=z_f m_b$ of the
internal quark, we parameterize the convolution as  
\begin{align}
\int_0^1 du \; h_9(u;z_f)  \; \phi_{M_2}(u)
\;=\;
H_{9,0}(z_f)
+ H_{9,1}(z_f) \,a_1^{M_2}
+ H_{9,2}(z_f) \,a_2^{M_2}.
\label{eq:conv:h9}
\end{align}
For massless quarks we may perform the convolution integral
analytically, 
\begin{align}
\int_0^1 du \; h_9(u;0)  \; \phi_{M_2}(u)
\;&=\;
\frac{493}{18} - \frac{2\pi^2}{3} - \left( \frac{40}{3} + 2\pi^2
    \right) a_1^{M_2}+ \left( \frac{8059}{600} - \pi^2 \right)
    a_2^{M_2},
\end{align}
whereas we obtain numerical results for massive internal quarks. In
Table~\ref{tab:convolutions} we summarize the contributions from closed
fermion loops for massless quarks ($z_q=0$), for a $b$-quark ($z_b=1$)
and for a charm quark ($z_c \in [0.25,0.35]$).  

\begin{table}[b!]
\centerline{
\parbox{13cm}{\setlength{\doublerulesep}{0.1mm}
\centerline{\begin{tabular}{|c||c||c|c|c|c|c||c|} \hline
\hspace*{1cm}&\hspace*{1cm}&\hspace*{1cm}&\hspace*{1cm}&
\hspace*{1cm}&\hspace*{1cm}&\hspace*{1cm}&\hspace*{1cm}
\\[-0.7em]
$z_f$ & $0$ & $0.25$ & $0.275$ & $0.3$ & $0.325$ &
$0.35$ & $1$
\\[0.3em]
\hline\hline\hline&&&&&&&
\\[-0.7em]
$H_{9,0}$ &
$20.81$ & $17.12$ & $16.43$ &
$15.72$ & $14.99$ & $14.26$ &
$-3.62$
\\[0.3em]
\hline&&&&&&&
\\[-0.7em]
$H_{9,1}$ &
$-33.07$ & $-13.28$ & $-12.37$ &
$-11.54$ & $-10.77$ & $-10.07$ &
$-0.68$
\\[0.3em]
\hline&&&&&&&
\\[-0.7em]
$H_{9,2}$ &
$3.56$ & $2.08$ & $1.94$ &
$1.81$ & $1.68$ & $1.57$ &
$0.01$
\\[0.3em]
\hline
\end{tabular}}
\vspace{4mm} \caption{\label{tab:convolutions}\small \textit{Fermionic
    contribution in the notation of (\ref{eq:conv:h9}). The first column
    refers to massless quarks, the last column to the $b$-quark and the
    other columns to the charm quark for different physical values of
    $z_c=m_c/m_b$.}}}} 
\end{table}

We illustrate the relative importance of the individual contributions
setting $\mu=m_b$ and $z_c=m_c/m_b=0.3$, which yields (with $C_F=4/3$,
$C_A=3$, $n_f=5$)  
\begin{align}
V^{(1)} &=
    ( -22.500 - 9.425\, i )
    + ( 5.500 - 9.425\, i ) \,a_1^{M_2}
    + (- 1.050 ) \,a_2^{M_2},\no \\
V_1^{(2)} &=
    ( -178.38 - 349.44\, i )
    + ( 660.59 - 119.36\, i ) \,a_1^{M_2}
    + ( -85.40 - 62.63\, i ) \,a_2^{M_2},\no \\
V_2^{(2)} &=
    ( 322.19 + 320.94\, i )
    + ( -212.97 + 154.41\, i ) \,a_1^{M_2}
    + ( 3.81 - 34.06\, i ) \,a_2^{M_2}.
\end{align}
We find relatively large coefficients for the NNLO terms and expect only
a minor impact of the higher Gegenbauer moments in the symmetric case
with $a_1^{M_2}=0$.  

We conclude with a remark concerning the large $\beta_0$-limit that has
been considered in~\cite{betazero}. In this approximation we get 
\begin{align}
V_1^{(2)} \big|_{\beta_0} &\simeq
(-239.31 -264.94i) + (380.33 - 252.90i) a_1^{M_2} + (-40.96 - 21.68i)
a_2^{M_2},
\end{align}
whereas the contribution from $V_2^{(2)}$ is completely missed. As a
consequence the NNLO contribution to $\alpha_1$ is substantially
underestimated in this approximation, whereas the one to $\alpha_2$
deviates from the full NNLO result between $\sim15\%$ for the imaginary
part and $\sim40\%$ for the real part. This illustrates the importance
of performing exact 2-loop calculations.


\section{Numerical analysis}

\label{sec:numerics}

We conclude with a brief analysis of the numerical impact of the
considered NNLO corrections. As a phenomenological analysis of hadronic
$B$ decays is beyond the scope of the present paper, we focus on the
perturbative structure of the topological tree amplitudes and discuss
their remnant uncertainties. In particular, we now combine our results
with the NNLO corrections from 1-loop spectator scattering that have
been worked out in~\cite{SpecScat:NLO:tree}.  

\subsection{Implementation of spectator scattering} 

In contrast to the vertex corrections considered in this work, the
spectator scattering term is sensitive to two perturbative scales: the
hard scale $\mu_h\sim m_b$ and a dynamically generated intermediate
(hard-collinear) scale $\mu_{hc}\sim (\LQCD m_b)^{1/2}$. The hard
scattering kernels from spectator scattering therefore factorize further
into coefficient functions $H_i^{II}$, encoding the hard effects, and a
universal hard-collinear jet-function $J_{||}$. Renormalization group
techniques can be used to resum parametrically large logarithms of the
form $\ln m_b/\LQCD$ in terms of an evolution kernel
${\cal{U}}_{||}$. Following the first paper of~\cite{SpecScat:NLO:tree},
we implement the spectator scattering contribution to the topological
tree amplitudes as\footnote{One should keep in mind that the Wilson
  coefficients in the spectator scattering term refer to a different
  operator basis than the one used in the current work (namely the
  Fierz-symmetric \emph{traditional basis} that we denoted by a tilde
  in~\cite{GB:Im}).}    
\begin{align}
& C_{i}(\mu) \;\,
  T_{i}^{II}(\mu) \otimes
  [\hat{f}_{B} \phi_B](\mu) \otimes
  \phi_{M_1}(\mu) \otimes
  \phi_{M_2}(\mu) \no \\
& \rightarrow \;
  C_{i}(\mu_h) \;\,
  H_{i}^{II}(\mu_h) \otimes
  {\cal{U}}_{||}(\mu_h,\mu_{hc}) \otimes
  J_{||}(\mu_{hc}) \otimes
  [\hat{f}_{B} \phi_B](\mu_{hc}) \otimes
  \phi_{M_1}(\mu_{hc}) \otimes
  \phi_{M_2}(\mu_h).
\label{eq:resummation}
\end{align}
Since the spectator scattering starts at $\calO(\as)$, the resummation
is required here in the next-to-leading-logarithmic (NLL)
approximation. Unfortunately, a complete NLL resummation is not possible
since the evolution kernel ${\cal{U}}_{||}$ is known in the
leading-logarithmic (LL) approximation only~\cite{Hill:2004if}.  

We therefore proceed along the lines of our earlier
analysis~\cite{GB:Im}, where we worked in the LL approximation which is
consistent for the imaginary parts that are of $\calO(\as^2)$. According
to this, we implement the LL evolution of the HQET decay constant and
the Gegenbauer moments to evolve the hadronic parameters from their
input scales to the ones required in (\ref{eq:resummation}). The $B$
meson distribution amplitude is modeled according to~\cite{Lee:2005gza},
which implies $\lambda_B(1\gev)=(0.48\pm0.12)\gev$ and, for the first
two logarithmic moments, $\sigma_1(1\gev)=1.6\pm0.2$ and
$\sigma_2(1\gev)=3.3\pm0.8$. The 1-loop matching corrections to the hard
functions $H_i^{II}$~\cite{SpecScat:NLO:tree} and the jet function
$J_{||}$~\cite{Hill:2004if,Becher:2004kk} are implemented neglecting
crossed terms of $\calO(\as^3)$. We finally adopt the BBNS model
from~\cite{BBNS} to estimate the size of power corrections to the
factorization formula.  

In the spectator scattering term we compute the Wilson coefficients from
the effective weak Hamiltonian in the NLL approximation with 2-loop
running coupling constant. Quantities referring to the hard scale are
evaluated in a theory with $n_f=5$ flavours and those referring to the
hard-collinear scale with $n_f=4$. 

\subsection{Tree amplitudes in NNLO}

We finally evaluate the topological tree amplitudes for the $B\to\pi\pi$
channels using the input parameters from our earlier
analysis~\cite{GB:Im} and computing the Wilson coefficients in the
vertex corrections in the next-to-next-to-leading-logarithmic (NNLL)
approximation~\cite{Gorbahn:2004my,Bobeth:1999mk} with 3-loop running
coupling constant~\cite{Tarasov:1980au} and
$\Lambda_\text{\tiny\MSbar}^{(5)}=205$~MeV. Under these specifications
the NNLO prediction of the topological tree amplitudes
becomes\footnote{The numbers for the imaginary parts differ slightly
  from those of~\cite{GB:Im}, since we now evaluate the Wilson
  coefficients throughout in the NNLL approximation.}    
\begin{align}
\alpha_1(\pi\pi)& \,=\,~~ 
                        1.008 \big{|}_{V^{(0)}}
         + \big[ 0.022 + 0.009i \big]_{V^{(1)}}
         + \big[ 0.024 + 0.026i \big]_{V^{(2)}}
\no\\
&\hspace{4.5mm}
                      - 0.012 \big{|}_{S^{(1)}}
         - \big[ 0.014 + 0.011i \big]_{S^{(2)}}
                      - 0.007 \big{|}_{P}
\no\\
& \,=\,~~ 
         1.019^{+0.017}_{-0.021} + (0.025^{+0.019}_{-0.015})i,
\no\\
\alpha_2(\pi\pi)& \,=\,~~   
                        0.224 \big{|}_{V^{(0)}}
         - \big[ 0.174 + 0.075i \big]_{V^{(1)}}
         - \big[ 0.030 + 0.048i \big]_{V^{(2)}}
\no\\
&\hspace{4.5mm}
                      + 0.075 \big{|}_{S^{(1)}}
         + \big[ 0.032 + 0.019i \big]_{S^{(2)}}
                      + 0.045 \big{|}_{P}
\no\\
& \,=\,~~  
         0.173^{+0.088}_{-0.073} - (0.103^{+0.051}_{-0.054})i.
\label{eq:NNLOresult}
\end{align} 
Here we disentangled the contributions of the various terms in the
factorization formula, namely the tree level result $V^{(0)}$ (''naive
factorization''), NLO (1-loop) vertex corrections $V^{(1)}$, NNLO
(2-loop) vertex corrections $V^{(2)}$, NLO (tree level) spectator
scattering $S^{(1)}$, NNLO (1-loop) spectator scattering $S^{(2)}$ and
the modelled power corrections $P$.  

\begin{table}[b!]
\centerline{
\parbox{13cm}{\setlength{\doublerulesep}{0.1mm}
\centerline{\begin{tabular}{|c||c|c||c|c|c|c||c|}\hline
\hspace*{1.2cm}&\hspace*{1.2cm}&\hspace*{1.2cm}&\hspace*{1.2cm}&\hspace*{1.2cm}&\hspace*{1.2cm}&\hspace*{1.2cm}&\hspace*{1.2cm}
\\[-0.7em] 
&$\mu_h$ & $\mu_{hc}$ &  $f_B$ & $F_+^{B\pi}$& $\lambda_B$ & $a_2^\pi$
& $X_H$\\[0.3em]
\hline\hline\hline&&&&&&& \\[-0.7em]
$\text{Re}(\alpha_1)$ &        
\large${}^{+0.008}_{-0.011}$ & 
\large${}^{+0.006}_{-0.007}$ & 
\large${}^{+0.003}_{-0.003}$ & 
\large${}^{+0.006}_{-0.008}$ & 
\large${}^{+0.006}_{-0.009}$ & 
\large${}^{+0.007}_{-0.008}$ & 
\large${}^{+0.007}_{-0.007}$   
\\[0.4em]
\hline&&&&&&& \\[-0.7em]
$\text{Im}(\alpha_1)$ &        
\large${}^{+0.017}_{-0.011}$ & 
\large${}^{+0.002}_{-0.003}$ & 
\large${}^{+0.001}_{-0.001}$ & 
\large${}^{+0.002}_{-0.003}$ & 
\large${}^{+0.002}_{-0.003}$ & 
\large${}^{+0.004}_{-0.004}$ & 
\large${}^{+0.007}_{-0.007}$   
\\[0.4em]
\hline\hline\hline&&&&&&& \\[-0.7em]
$\text{Re}(\alpha_2)$ &        
\large${}^{+0.016}_{-0.008}$ & 
\large${}^{+0.026}_{-0.023}$ & 
\large${}^{+0.014}_{-0.014}$ & 
\large${}^{+0.038}_{-0.025}$ & 
\large${}^{+0.039}_{-0.026}$ & 
\large${}^{+0.038}_{-0.033}$ & 
\large${}^{+0.045}_{-0.045}$   
\\[0.4em]
\hline&&&&&&& \\[-0.7em]
$\text{Im}(\alpha_2)$ &        
\large${}^{+0.019}_{-0.028}$ & 
\large${}^{+0.005}_{-0.004}$ & 
\large${}^{+0.002}_{-0.002}$ & 
\large${}^{+0.005}_{-0.003}$ & 
\large${}^{+0.005}_{-0.003}$ & 
\large${}^{+0.007}_{-0.006}$ & 
\large${}^{+0.045}_{-0.045}$   
\\[0.4em]
\hline
\end{tabular}}
\vspace{4mm} \caption{\label{tab:uns}\small \textit{Dominant
    uncertainties of our final predictions for the colour-allowed tree
    amplitudes $\alpha_1(\pi\pi)$ and the colour-suppressed tree
    amplitude $\alpha_2(\pi\pi)$ from scale variations, hadronic input
    parameters and modelled power corrections.}}}}  
\end{table}

The new contributions from this work consist in the real parts of the
terms denoted by $V^{(2)}$. For the colour-allowed amplitude
$\alpha_1(\pi\pi)$, this correction is slightly larger than the $\as$
terms due to an numerical enhancement from the Wilson coefficients in
the effective Hamiltonian\footnote{We remark that a similar enhancement
  is unlikely to exist at even higher order of the perturbative
  expansion, since the NNLO expressions already reveal the full
  complexity.}. On the other hand, the colour-suppressed amplitude
$\alpha_2(\pi\pi)$ receives a moderate correction. In particular, we do
not find an enhancement of the phenomenologically interesting ratio
$|\alpha_2/\alpha_1|$ from the perturbative calculation.      
  
\begin{figure}[b!]
\centerline{\parbox{16cm}{
\centerline{
 \psfrag{muh}{$\mu_h$}
 \psfrag{Rea1}{$\text{Re}(\alpha_1)_V$}
 \includegraphics[width=7cm]{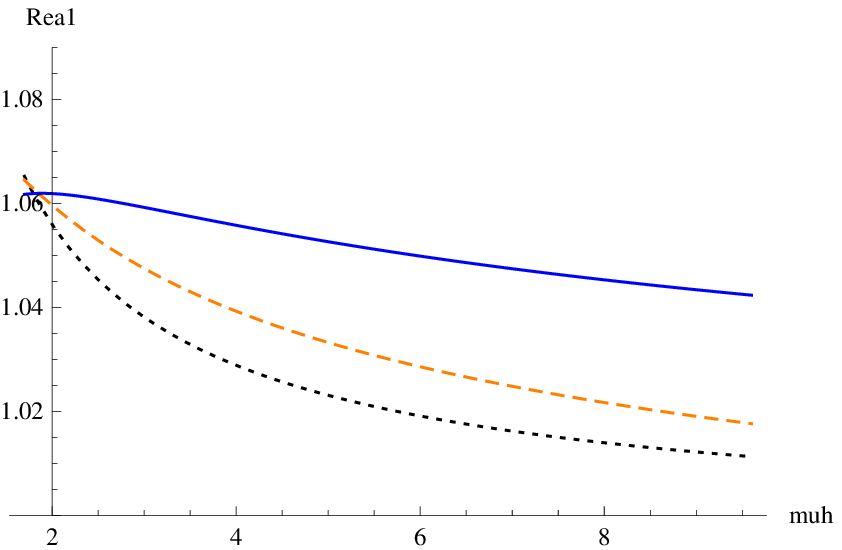}
\hspace{10mm}
 \psfrag{muh}{$\mu_h$}
 \psfrag{Ima1}{$\text{Im}(\alpha_1)_V$}
 \includegraphics[width=7cm]{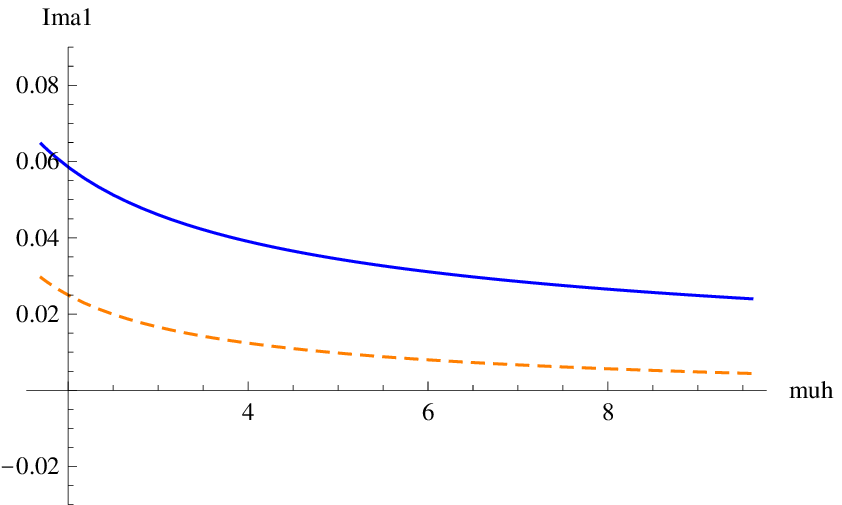}}
\vspace{10mm}
\centerline{
 \psfrag{muh}{$\mu_h$}
 \psfrag{Rea2}{$\text{Re}(\alpha_2)_V$}
 \includegraphics[width=7cm]{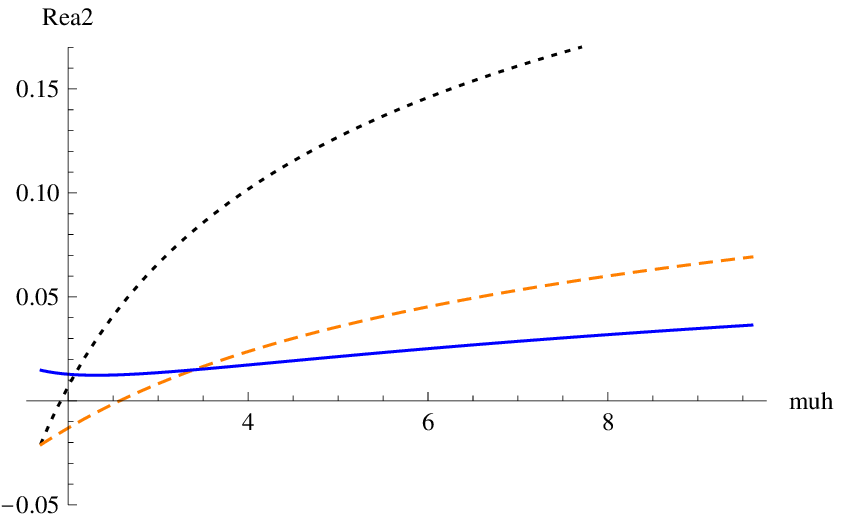}
\hspace{10mm}
 \psfrag{muh}{$\mu_h$}
 \psfrag{Ima2}{$\text{Im}(\alpha_2)_V$}
 \includegraphics[width=7cm]{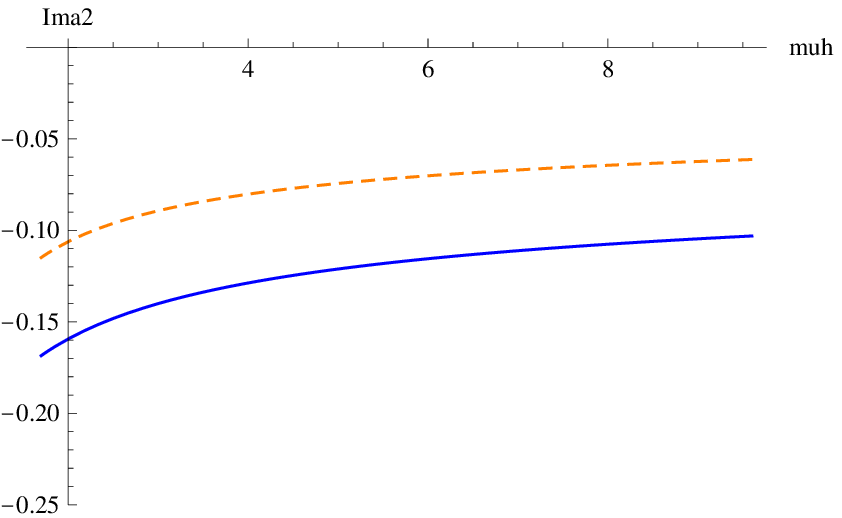}}
\caption{\label{fig:scale} \small \textit{Dependence of the tree
    amplitudes $\alpha_i(\pi\pi)$ as a function of the hard scale
    $\mu_h$ (vertex corrections only). The dotted (black) lines refer to
    LO, the dashed (orange/light gray) lines to NLO and the solid
    (blue/dark gray) lines to NNLO.}}}} 
\end{figure}

In Table~\ref{tab:uns} we list the uncertainties of our NNLO predictions
stemming from scale variations, hadronic input parameters and the
modelled power corrections. The values of the first two columns follow
from varying the perturbative scales independently in the ranges
$\mu_h=4.8^{+4.8}_{-2.4}~\gev$ and $\mu_{hc}=1.5^{+0.9}_{-0.5}~\gev$. As
the dependence on the hard scale tends to cancel between vertex
corrections and spectator scattering, we vary both contributions
independently and take the larger interval (from the vertex corrections)
as our estimate for higher order perturbative corrections. The scale
dependence of the vertex corrections is also illustrated in
Figure~\ref{fig:scale}, where we read off that it gets substantially
reduced for the real parts at NNLO, whereas the reduction is less
pronounced for the imaginary parts.

For our final error estimate in (\ref{eq:NNLOresult}) we added the
individual uncertainties from Table~\ref{tab:uns} in quadrature. Whereas
the colour-allowed amplitude $\alpha_1(\pi\pi)$ can be computed
precisely in the factorization framework, the situation is less
fortunate for the colour-suppressed amplitude $\alpha_2(\pi\pi)$. Due to
large cancellations between the vertex corrections, the
colour-suppressed amplitude becomes particularly sensitive to the
spectator scattering contribution and is therefore subject to rather
large uncertainties related mainly to our restricted knowledge of the
hadronic input parameters.


\section{Conclusion}

\label{sec:conclusion}

We computed the real parts of the 2-loop vertex corrections for
charmless hadronic $B$ meson decays, completing the NNLO calculation of
the topological tree amplitudes in the QCD factorization framework. We
in particular showed how to compute the colour-suppressed tree amplitude
without making use of Fierz-symmetry arguments and found that the
hard-scattering kernels are free of IR-singularities and the resulting
convolutions with the light-cone distribution amplitude of the emitted
light meson are finite, which demonstrates factorization at the 2-loop
order.    

The numerical impact of the considered corrections was found to be
moderate, although they can be of similar size as the NLO
corrections. The scale dependence of the real parts of the topological
tree amplitudes is significantly reduced at NNLO, which allows for a
precise determination of the colour-allowed amplitude $\alpha_1$. In
contrast to this, it remains difficult to compute the colour-suppressed
amplitude $\alpha_2$ in the factorization framework, since it is subject
to substantial uncertainties from hadronic input parameters and
potential $1/m_b$ corrections. In particular, we do not find an
enhancement of the phenomenologically important ratio
$|\alpha_2/\alpha_1|$ from the perturbative calculation.

\subsection*{Acknowledgements}


We are grateful to Gerhard Buchalla for interesting discussions and
helpful comments on the manuscript. This work was supported by the DFG
Sonderforschungsbereich/Trans\-regio 9.

\begin{appendix}

\section{Auxiliary coefficient functions}
\label{app:coefffuncs}

In the calculation of the colour-allowed tree amplitude, the NLO kernels
have been given in (\ref{eq:T1:a1}) in terms of the coefficient
functions 
\begin{align}
t_0(u) &=  4 \Li_2(u) - \ln^2 u +2 \ln u \ln \ubar + \ln^2 \ubar
+(2-3u) \Big( \frac{\ln u}{\ubar} - \frac{\ln \ubar}{u} \Big)
-\frac{\pi^2}{3} -22, \no
\\
t_1(u) &= -2 \Li_3(u) -2 \S_{1,2}(u) -2 \ln \ubar \, \Li_2(u) + \ln^3 u
-2 \ln^2 u \ln \ubar + \ln u \ln^2 \ubar - \ln^3 \ubar \no \\
& \quad \,+ \frac{2-3u^2}{u\ubar} \Li_2(u) - \frac{2-3u}{\ubar}
\Big(\ln^2 u - \ln u \ln \ubar \Big) + \frac{6-11u +2\ubar \pi^2}{\ubar}
\ln u \no \\
& \quad \,+\frac{4-3u}{2u} \ln^2 \ubar - \frac{18-33u+5u \pi^2}{3u}\ln
\ubar  +\frac{(7-6u)\pi^2}{6\ubar} +2 \zeta_3 -52, \no
\\
t_2(u) &= 10 \Li_4(u) -8 \S_{2,2}(u) +10 \S_{1,3}(u) -8 \ln \ubar \,
\Li_3(u) +10 \ln \ubar \, \S_{1,2}(u) - \frac{7}{12} \ln^4 u  \no \\
& \quad \, +5 \ln^2 \ubar \, \Li_2(u) + \frac43 \ln^3 u \ln \ubar -\ln^2
u \ln^2 \ubar +\frac13 \ln u \ln^3 \ubar +\frac{7}{12} \ln^4\ubar \no \\
& \quad \, + \frac{2-6u+6u^2}{u \ubar} \Li_3(u)
-\frac{4-6u+3u^2}{u\ubar} \Big( \S_{1,2}(u) +\ln \ubar\,\Li_2(u) \Big)
-\frac{8-3u}{6u} \ln^3 \ubar \no\\
& \quad \, + \frac{2-3u}{6\ubar} \Big( 4 \ln^3 u - 6\ln^2 u \ln\ubar +3
\ln u \ln^2 \ubar \Big) - \frac{60(1-2u)+17\ubar\pi^2}{12\ubar}\ln^2 u
\no \\
& \quad \, + \frac{3(6-4u-7u^2)+u\ubar \pi^2}{3u\ubar} \Li_2(u)
+\frac{24-54u+5\ubar \pi^2}{6\ubar} \ln u \ln \ubar
+\frac{(29-24u)\pi^2}{6\ubar} \no \\
& \quad \, + \frac{6(12-13u)+7u\pi^2}{12u} \ln^2 \ubar
+\frac{24(7-13u)+(10-15u)\pi^2}{12\ubar} \ln u -
\frac{23\pi^4}{180}\no\\
& \quad \, - \frac{24\ubar(7-13u)+(2+23u-27u^2)\pi^2+24u\ubar
  \zeta_3}{12u\ubar} \ln \ubar + \frac{10-11u}{\ubar} \zeta_3-112, \no
\\
t_{E,0}(u) &=-\frac{1-2u}{2} \Big( \frac{\ln u}{\ubar} - \frac{\ln
  \ubar}{u} \Big) + \frac{16}{3}, \no
\\
t_{E,1}(u) &= -\frac{1-2u}{2u\ubar} \Li_2(u) + \frac{1-3u}{4\ubar} \ln^2
u + \frac{u}{2\ubar} \ln u \ln \ubar -\frac{2-3u}{4u} \ln^2 \ubar \no \\
& \quad \, -\frac{4(1-2u)}{3} \Big( \frac{\ln u}{\ubar} - \frac{\ln
  \ubar}{u} \Big) - \frac{(6-5u)\pi^2}{12\ubar}+12
\end{align}
and the convolutions of the NLO kernels with the wave function
corrections, cf.~(\ref{eq:ConT1:a1}) and (\ref{eq:ConT1E:a1}), involve 
\begin{align}
t_3(u) &= 4 \Li_3(u) + 4 \S_{1,2}(u) -4 \ln u \, \Li_2(u) +\frac23 \ln^3
u -2 \ln^2 u \ln \ubar -\frac23 \ln^3 \ubar - \frac{\Li_2(u)}{u\ubar}
\no \\
& \quad - \frac{1-3u}{2u\ubar} \Big( u \ln^2u +2 \ubar \ln u \ln \ubar -
\ubar \ln^2 \ubar \Big)\! - \frac{3}{2u} \ln \ubar +
\frac{(4-3u)\pi^2}{6\ubar}-\frac{15}{2}-4 \zeta_3 ,\no
\\
t_4(u) &= 12 \Li_4(u) -20 \S_{2,2}(u) +12 \S_{1,3}(u) -8 \Big(\ln u +
\ln \ubar \Big) \Li_3(u)  +12 \ln u \,\S_{1,2}(u) \no \\
& \quad +4 \ln \ubar \, \S_{1,2}(u) + \! \Big(4\ln^2 u +4 \ln u \ln
\ubar +2 \ln^2 \ubar \Big) \Li_2(u)  - \frac34 \ln^4 u +\frac73 \ln^3 u
\ln \ubar \no \\
& \quad -\frac12 \ln^2 u \ln^2 \ubar -\frac13 \ln u \ln^3 \ubar +\frac34
\ln^4 \ubar - \frac{4-11u+3u^2}{u\ubar} \Li_3(u) + \frac{5-12u}{6\ubar}
\ln^3 u\no \\
& \quad + \frac{1+u-3u^2}{u\ubar} \S_{1,2}(u) +
\frac{2-10u+6u^2}{u\ubar} \ln u \, \Li_2(u) - \frac{1-5u+3u^2}{u\ubar}
\ln \ubar \, \Li_2(u) \no \\
 & \quad + \frac{2-10u+9u^2}{2u\ubar} \ln^2 u \ln \ubar -
 \frac{1-2u}{2u\ubar} \ln u \ln^2 \ubar - \frac{5-6u}{6u} \ln^3 \ubar
 \no \\
& \quad - \frac{18-24u+15u^2-10u\ubar \pi^2}{3u\ubar} \Li_2(u) -
\frac{16-27u+4\ubar\pi^2}{4\ubar} \ln^2 u \no \\
& \quad - \frac{6-36u+27u^2-4u\ubar\pi^2}{2u\ubar} \ln u \ln \ubar +
\frac{3(14-17u) +8u\pi^2}{12u} \ln^2 \ubar  \no \\
& \quad + \frac{8-15u-4\pi^2-48\ubar \zeta_3}{4\ubar} \ln u  +
\frac{3(2-3u)\zeta_3}{\ubar}  - \frac{23\pi^4}{60} +
\frac{(23-17u)\pi^2}{12\ubar}  \no \\
& \quad -
\frac{81-126u+45u^2-(14-22u+6u^2)\pi^2-192u\ubar\zeta_3}{12u\ubar} \ln
\ubar - \frac{137}{4}, \no
\\
t_{E,2}(u) &= -\frac{6(1-2u)}{u\ubar}\Li_2(u) -\frac{6}{u} \ln u \ln
\ubar -6\ln u-6\ln \ubar-\frac{\pi^2}{\ubar}+50.
\end{align}
In the calculation of the colour-suppressed tree amplitude, the NLO
kernels in (\ref{eq:T12:a2}) contain the coefficient functions 
\begin{align}
\hat{t}_1(u) &=
\frac{u}{\ubar} \ln u - \ln \ubar + 8 + i \pi, \no \\
\hat{t}_2(u) &=
\frac{u}{\ubar} \Big( \Li_2(u) - \ln^2 u + \ln u \ln \ubar + 4 \ln u
    \Big) + \frac12 \ln^2 \ubar - 4 \ln \ubar - \frac{(3 -
      2u)\pi^2}{6\ubar} + 20
\no\\
& \quad
+ i \pi \Big(4 - \ln \ubar \Big),
\no \\%
\hat{t}_{E,0}(u) &=
\frac{\ubar}{u} \ln \ubar - \ln u  + 6 + i \pi,
\no \\%
\hat{t}_{E,1}(u) &=
- \frac{\ubar}{u} \Big( \Li_2(u) + \ln^2 \ubar - 3 \ln \ubar \Big)
+ \frac12 \ln^2 u - 3 \ln u + 14 - \frac{\pi^2}{2}
+ i \pi \Big(3 - \ln u \Big)
\end{align}
and the convolutions with the NLO kernels, cf.~(\ref{eq:ConT12:a2}) and
(\ref{eq:ConT12EF:a2}), give rise to 
\begin{align}
\hat{t}_3(u) &=
\frac{\pi^2}{3} - 5 + \frac{u}{\ubar} \ln^2 u + 2 \ln u \ln \ubar
- \ln^2\ubar - \frac{\ln \ubar}{u},
\no \\
\hat{t}_{E,2}(u) &=
\frac{1}{\ubar} \Big( 6 \Li_2(u) + 3 u \ln u - \pi^2 \Big)
- \frac{3(1 + u)}{u} \ln \ubar + 27 + 3 i \pi,
\no \\
\hat{t}_{F,0}(u) &=
\frac{1}{\ubar} \Big( 2(1 + 2u) \Li_2(u) - u \ln^2 u - 2\ubar \ln u \ln 
\ubar + 3u \ln u - (2 + u) \frac{\pi^2}{3} \Big)
\no \\
&\quad
+ \frac{\ubar}{u} \ln^2 \ubar - \frac{3(1 + u)}{u} \ln \ubar
+ 29 + i \pi \Big( 3 - \frac{2u}{\ubar}  \ln u + \frac{2\ubar}{u} \ln
\ubar \Big).
\end{align}

\section{Calculation of 2-loop counterterms $\hat{\Delta}_{F,i}^{(2)}$}
\label{app:Delta2}

We present the calculation of the 2-loop counterterms
$\hat{\Delta}_{F,i}^{(2)}$, that are required in the NNLO calculation of
the colour-suppressed tree amplitude as described in
Section~\ref{sec:IR}. The counterterms receive three contributions 
\begin{align}
\hat{\Delta}_{F,i}^{(2)} &= \hat{T}_{i,F}^{(0)} \otimes
\bigg\{
\hat{F}_{F}^{(2)} \,\hat{\Phi}_F^{(0)}
+ \hat{F}_{F}^{(1)} \,\hat{\Phi}_{F}^{(1)}
+ \hat{F}_{F}^{(0)} \,\hat{\Phi}_{F}^{(2)}
\bigg\}_\text{\footnotesize ct}
\end{align}
where $\otimes$ represents a convolution and ''ct'' refers to the
counterterm contributions of the form factor and the wave function
corrections. Notice that the wave function corrections actually
correspond to local corrections to the decay constant, as a consequence
of the fact that the tree level kernels $\hat{T}_{i,F}^{(0)}$ are
constant (cf.~also (\ref{eq:ct:1loop}) and (\ref{eq:Delta1})).  

\begin{figure}[t!]
\centerline{\parbox{13cm}{
\centerline{\includegraphics[height=2cm]{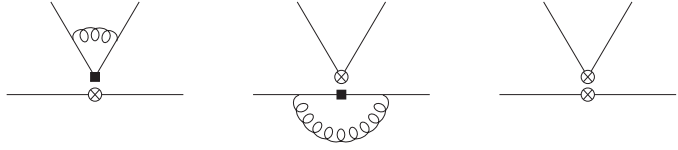}}
\caption{\label{fig:F1Phi1}  \small \textit{Diagrams that contribute to
    the mixed contribution $\hat{F}_{F}^{(1)}\,\hat{\Phi}_F^{(1)}$. The
    symbol $\otimes$ in the lower (upper) line refers to an insertion of
    the 1-loop counter\-term from the form factor (wave function)
    correction of the operator $\tilde{\calO}_F$.}}}} 
\end{figure}

\begin{figure}[b!]
\centerline{\parbox{13cm}{
\centerline{\includegraphics[height=2cm]{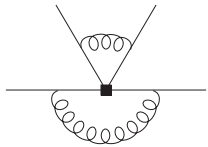}}
\caption{\label{fig:FF}  \small \textit{The UV-divergences of this
    2-loop diagram contribute to $\delta_2$.}}}} 
\end{figure}

\begin{figure}[t!]
\centerline{\parbox{13cm}{
\centerline{\includegraphics[height=2.2cm]{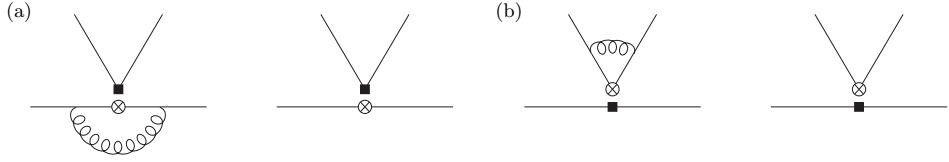}}
\caption{\label{fig:F2Phi0}  \small \textit{Diagrams that contribute to
    $\hat{F}_{F}^{(2)}\,\hat{\Phi}_F^{(0)}$~(a) and
    $\hat{F}_{F}^{(0)}\,\hat{\Phi}_F^{(2)}$~(b).}}}} 
\end{figure}

We first consider the mixed term
$\hat{F}_{F}^{(1)}\,\hat{\Phi}_F^{(1)}$,  which involves the calculation
of the diagrams from Figure~\ref{fig:F1Phi1}. The first diagram vanishes
due to a scaleless loop integral and the second diagram yields 
\begin{align}
\delta_1 &= - C_F^2
\left(\frac{e^{\gamma_E}\mu^2}{m_b^2} \right)^\eps \Gamma(\eps)
\bigg\{
\bigg( \frac{1-\eps+2\eps^2}{\eps(1-2\eps)}
+ \frac{6(1+\eps)}{(1-\eps)^2} \bigg)
\hat{F}^{(0)}  \,\hat{\Phi}^{(0)}
+ \frac{6(1+\eps)}{(1-\eps)^2} \;
\hat{F}_F^{(0)}\,\hat{\Phi}_F^{(0)}
\no\\
&\hspace{3.8cm}
+ \frac{1-6\eps+16\eps^2-14\eps^3}{4\eps^2(1-2\eps)(1-\eps)^2}
\; \hat{F}_E^{(0)} \,\hat{\Phi}_E^{(0)}
+ \frac{1}{16\eps(1-\eps)^2}
\; \hat{F}_{E'}^{(0)} \,\hat{\Phi}_{E'}^{(0)}
\bigg\}.
\label{eq:delta1}
\end{align}
We are left with the 2-loop counterterm from the last diagram of
Figure~\ref{fig:F1Phi1}, which requires the calculation of the
UV-divergences of the 2-loop diagram from Figure~\ref{fig:FF}. For this
it is convenient to apply the method proposed in~\cite{IRrearrange}
(sometimes called \emph{IR-rearrangement}), which allows to set all
masses and external momenta to zero. The calculation then reduces to the
evaluation of 2-loop tadpole integrals, which depend on a single mass
scale (an artificial scale that has been introduced to separate UV- and
IR-divergences). Computing the 1-loop counterterms with the same
prescription and accounting for the wave-function renormalization, we
get 
\begin{align}
\delta_2 &= C_F^2 \bigg\{
\bigg( \frac{6}{\eps}+ 5\bigg)\hat{F}^{(0)}  \,\hat{\Phi}^{(0)}
- \bigg( \frac{3}{4\eps^2} - \frac{1}{\eps}\bigg)
\hat{F}_E^{(0)} \,\hat{\Phi}_E^{(0)}
+ \frac{1}{16\eps^2} \hat{F}_{E'}^{(0)} \,\hat{\Phi}_{E'}^{(0)}
+ \frac{6}{\eps} \hat{F}_F^{(0)}\,\hat{\Phi}_F^{(0)}
\bigg\}.
\label{eq:delta2}
\end{align}
Next we compute the form factor correction $\hat{F}_{F}^{(2)}
\,\hat{\Phi}_F^{(0)}$ (the corresponding diagrams are shown in
Figure~\ref{fig:F2Phi0}a). The first diagram gives again the
contribution $\delta_1$ from~(\ref{eq:delta1}). On the other hand the
computation of the 2-loop counterterm from the second diagram of
Figure~\ref{fig:F2Phi0}a is rather involved. It requires the calculation
of the UV-divergences of a couple of 2-loop diagrams (shown e.g.~in
Figure~1 of~\cite{GB:Inclusive}) and the corresponding 1-loop
counterterms. Proceeding as before with the method of
\emph{IR-rearrangement} and accounting for the 2-loop wave-function
renormalization in the \MSbar-scheme~\cite{Egorian:1978zx}, 
\begin{align}
Z_{2,b}^{(2)} &= Z_{2,q}^{(2)} =
C_F \bigg\{ \bigg( \frac{1}{2} C_F + C_A \bigg) \frac{1}{\eps^2}
+ \bigg( \frac34 C_F - \frac{17}{4} C_A + \frac{1}{2}n_f
\bigg)\frac{1}{\eps} \bigg\},
\end{align}
yields the 2-loop form factor counterterm for the Fierz-evanescent
operator $\tilde{\calO}_F$\footnote{We performed this calculation for
  arbitrary bilinear quark currents, which allows us to perform several
  cross-checks. We in particular verified that the anomalous dimension
  of the vector current vanishes at the 2-loop level and reproduced the
  one of the scalar and the tensor current from~\cite{Tarrach:1980up}.} 
\begin{align}
\delta_3 &= C_F\bigg\{
\left[ \bigg( 3 C_F - \frac{11}{6} C_A + \frac13 n_f \bigg)
  \frac{1}{\eps}
+ \bigg(- \frac{17}{2} C_F + \frac{149}{36} C_A + \frac{5}{18} n_f
\bigg) 
  \right] \hat{F}^{(0)}  \,\hat{\Phi}^{(0)}
\no\\
&\quad
+\left[ \bigg( -\frac{3}{8} C_F - \frac{11}{24} C_A + \frac{1}{12} n_f
  \bigg) 
  \frac{1}{\eps^2}
+ \bigg( \frac{9}{16} C_F + \frac{53}{144} C_A - \frac{1}{72} n_f \bigg)
  \frac{1}{\eps}  \right] \hat{F}_E^{(0)} \,\hat{\Phi}_E^{(0)}
\no\\
&\quad
+ \left[ \frac{C_F}{32\eps^2}
+ \bigg(- \frac{5}{64} C_F + \frac{1}{32} C_A \bigg)  \frac{1}{\eps}
  \right] \hat{F}_{E'}^{(0)} \,\hat{\Phi}_{E'}^{(0)}
+\bigg( 3 C_F -\frac{11}{6} C_A + \frac13 n_f\bigg)  \frac{1}{\eps}
\hat{F}_F^{(0)}\,\hat{\Phi}_F^{(0)}
\bigg\}.
\label{eq:delta3}
\end{align}
We finally account for the wave function correction $\hat{F}_{F}^{(0)}
\,\hat{\Phi}_F^{(2)}$ from Figure~\ref{fig:F2Phi0}b. The first diagram
again vanishes due to a scaleless integral and the second diagram
yields, in a convolution with a constant kernel, again the contribution
$\delta_3$ from~(\ref{eq:delta3}).

To summarize, in terms of the individual contributions $\delta_i$ from
(\ref{eq:delta1}), (\ref{eq:delta2}) and (\ref{eq:delta3}), the 2-loop
counterterms required in the calculation of the colour-suppressed
amplitude become 
\begin{align}
\hat{\Delta}_{F,1}^{(2)} &=
C_F \hat{\Delta}_{F,2}^{(2)} =
\frac{C_F}{N_C}
\bigg\{ 2\delta_1 + \delta_2 + 2\delta_3 \bigg\}.
\end{align}

\section{NNLO hard-scattering kernels}
\label{app:kernels}

Our final expressions for the real parts of the NNLO vertex corrections
from (\ref{eq:V2}) involve the following set of hard-scattering kernels,
\begin{align}
&h_5(u) =
\bigg[
\frac{4(3 - 3u + 8u^2 - 2u^3)}{\ub^3} \Li_4(u)
- \frac{8(12 - 35u + 36u^2 - 14u^3 + 4u^4)}{u^3\ub^2} \S_{2,2}(u)
\no\\
&\quad
- 24 \ln u \,\Li_3(u)
- \frac{8(12 - 47u + 71u^2 - 48u^3 + 24u^4)}{u^3\ub^3}
\bigg( \ln \ub \, \Li_3(u) - \zeta_3 \ln u\bigg)
\no\\
&\quad
+ \frac{17 - 82 u^2 + 40 u^4 - 16u^6}{2u^3\ub^3} \Li_2(u)^2
+ 4 \ln^2 u \, \Li_2(u)
+ \frac23 \ln^4 u
- \frac43 \ln^3 u \ln \ub
\no\\
&\quad
- \frac{122 - 91 u^2 + 20u^4}{3u\ub^3} \ln u \, \ln \ub \, \Li_2(u)
- \frac{51 + 16u^2 + 268  u^4 + 4u^6}{12u^3\ub^3} \ln^2 u \ln^2 \ub
\no\\
&\quad
+ \frac{3 - 108 u + 351u^2 - 440 u^3 + 266 u^4 - 74u^5 + 17 u^6 - 16u^7
  + 4 
u^8}{u^3\ub^3} \Li_3(u)
\no\\
&\quad
- \frac{3 - 78 u + 109 u^2 - 43u^3 - 6 u^4}{u^3\ub} \ln u \, \Li_2(u)
- \frac{u(3 - 3u + 2u^2)}{6\ub} \ln^3 u
\no\\
&\quad
- \frac{3 - 54u + 69u^2 - 18u^3 - 6u^4 + 3u^5 - 2u^6}{2u^3\ub} \ln^2 u
\ln \ub
\no\\
&\quad
+ \frac{3 - 3u - 7u^2 - 3u^3}{u^3} \bigg( \Li_3(-u) - \ln u \, \Li_2(-u)
- \frac{\ln^2 u+\pi^2}{2}\ln(1+u) \bigg)
\no\\
&\quad
- \frac{(\ub-u)(6(5 - 66 u^2 + 60u^4 - 4u^6) + 4(19 - 91u^2 - 2u^4)
  \pi^2)}{24u^3\ub^3} \Li_2(u)
\no\\
&\quad
+ \bigg( \frac{42 - 57u - 22u^2 + 49u^3 - 4u^4}{4u\ub^2} - 2 \pi^2
\bigg) 
\ln^2 u
\no\\
&\quad
+ \bigg( \frac{273 - 535 u + 302 u^4 - 90u^5}{40u^2\ub^3}
+ \frac{(701 - 55 u + 2794 u^4 - 620u^5)\pi^2}{180u^2\ub^3} \bigg) \ln
u \ln \ub
\no\\
&\quad
- \bigg( \frac{167 - 302u}{4\ub} - \frac{(96 - 160u + 59u^2 + 9u^3 +
  3u^4 
  - 2 u^5)\pi^2}{6u^2\ub} \bigg) \ln u
\no\\
&\quad
+ \frac{(849 - 3456u + 4496u^2 - 2408u^3 + 2024u^4 - 984u^5 +
  328u^6)\pi^4}{720u^3\ub^3}
\no\\
&\quad
+ \frac{(5 - 262u + 938u^2 - 1582u^3 + 1366u^4 - 690u^5
  +230u^6)\pi^2}{48u^3\ub^3}
+ \frac{1507}{8}
\no\\
&\quad
- \frac{(3 - 110u + 358u^2 - 424u^3 + 32u^4 + 216u^5 -
  72u^6)\zeta_3}{2u^3\ub^3} 
+ (u\leftrightarrow\ubar)\bigg]
\no \\
&
+ \bigg[
- \frac{8(1 + 2u)}{\ub^3} \Li_4(u)
- \frac{8(12 - 35u + 36u^2 - 14u^3 + 4u^4)}{u^3\ub^2} \S_{2,2}(u)
\no \\
&\quad
- \frac{(\ub-u)(12 - 23u + 25u^2 - 4u^3 + 2u^4)}{u^3\ub^3}
\bigg( 8\ln \ub \,\Li_3(u)+ \ln^2 u \ln^2 \ub  - 8\zeta_3 \ln u \bigg)
\no \\
&\quad
- \frac{2(\ub-u)(3 - 5u + 5u^2)}{u^3\ub^3} \Li_2(u)^2
- \frac{3 - 72u + 85u^2 - 9u^3 + 18u^4}{u^3\ub} \ln u \, \Li_2(u)
\no \\
&\quad
- \frac{2(3 - 11u + 15u^2 - 8u^3 + 4u^4)}{u^3\ub^3} \bigg(\ln u \ln \ub 
- \frac{7 \pi^2}{6} \bigg) \Li_2(u)
\no \\
&\quad
+ \frac{3 - 102u + 293u^2 - 278u^3 + 36u^4 + 74u^5 - 43u^6 +
  14u^7}{u^3\ub^3} 
\Li_3(u)
\no \\
&\quad
- \frac{12 - 25u + 7u^2}{6\ub} \ln^3 u
- \frac{3 - 48u + 57u^2 - 6u^3 + 16u^4 - 7u^5}{2u^3\ub} \ln^2 u \ln \ub
\no \\
&\quad
+ \frac{3 + 3u + 5u^2 + 3u^3}{u^3} \bigg( \Li_3(-u) - \ln u \,\Li_2(-u) -
\frac{\ln^2 u+\pi^2}{2}\ln (1+u) \bigg)
\no \\
&\quad
+ \frac{21 - 62u - 29u^2 + 182u^3 - 91u^4}{2u^2\ub^2} \Li_2(u)
+ \frac{42 + u - 142u^2 + 91u^3}{4u\ub^2} \ln^2 u
\no \\
&\quad
+ \frac{(\ub-u)(3 u \ub(21 - 13u + 13u^2) + 4(33 - 61u + 65u^2 - 8u^3 +
4u^4)\pi^2)}{12u^3\ub^3} \ln u \ln \ub
\no \\
&\quad
- \frac{111u^2 - 2(96 - 112u - 29u^2 + 37u^3 - 7u^4)\pi^2}{12u^2\ub} \ln
u
\no \\
&\quad
+ \frac{(\ub-u)(48 - 83u + 85u^2 - 4u^3 + 2u^4)\pi^4}{90u^3\ub^3}
- \frac{(\ub-u)(147 - 74u + 74u^2)\pi^2}{24u^2\ub^2}
\no \\
&\quad
- \frac{(\ub-u)(3 - 110u + 96u^2 + 28u^3 - 14u^4)\zeta_3}{2u^3\ub^3}
- (u\leftrightarrow\ubar) \bigg],
\no \\
&h_6(u) =
\bigg[  \frac{327}{2} - \frac{3(1-2u)}{2\ubar} \ln^2 u +
\frac{3(1-2u^2)}{2u\ubar} \ln u \ln \ubar - \frac{3(13-24u)}{2\ubar} \ln
u
\no \\
&\quad
+ \frac{(1-2u^2)\pi^2}{4u\ubar} + (u\leftrightarrow\ubar)\bigg]
\no \\
&
+ \bigg[  8 \Li_3(u) -8 \ln u \, \Li_2(u) + \frac43 \ln^3 u -4 \ln^2 u
\ln \ubar - \frac{13-24u^2}{u\ubar} \Li_2(u)
\no \\
&\quad  + \frac{25-24u}{2\ubar} \ln^2 u + \frac{13}{\ubar} \ln u
\ln \ubar - \frac{9}{2\ubar} \ln u - \frac{11\pi^2}{6\ubar} -
(u\leftrightarrow\ubar) \bigg],
\no \\
&h_7(u) =
\bigg[
\frac{(1 + u)(3 - 4u + 3u^2)}{3u\ub^2} \bigg( 12\calH_1(u) +
\pi^2\ln(1+u)\bigg)
- 48\ln u\,\Li_3(u)
\no \\
&\quad
+ \frac{2u}{3\ub^3} \bigg( 24 \calH_2(u)- 2\pi^2 \Li_2(-u)\bigg)
+ \frac{4(6 - 9 u + 16 u^2 - 4u^3)}{\ub^3} \Li_4(u)
+ 8\ln^2 u\, \Li_2(u)
\no \\
&\quad
- \frac{4(52 - 152u + 156u^2 - 61u^3 + 18u^4 - u^5)}{u^3\ub^2}
\S_{2,2}(u)
+ \frac43\ln^4 u
- \frac83 \ln^3 u \ln \ub
\no \\
&\quad
- \frac{4(52 - 204u + 308u^2 - 209u^3 + 107u^4 - 3u^5 + u^6)}{u^3\ub^3}
\bigg( \ln \ub \, \Li_3(u) - \zeta_3 \ln u\bigg)
\no \\
&\quad
- \frac{13 - 54u + 88u^2 - 84u^3 + 82u^4 - 48u^5 + 16u^6}{u^3\ub^3}
\Li_2(u)^2
+ \frac{3 - 18u + 6u^2 - 4u^3}{6\ub} \ln^3 u
\no \\
&\quad
- \frac{(\ub - u)(1 - 2u + 2u^2)(13 - 2u + 2u^2)}{u^3\ub^3} \ln u
\,\ln\ub \,\Li_2(u)
\no \\
&\quad
- \frac{6 - 168u + 235u^2 - 107u^3 - 6u^4}{u^3\ub} \ln u \,\Li_2(u)
\no \\
&\quad
- \frac{52 - 204u + 308u^2 - 197u^3 + 71u^4 + 33u^5 -11u^6}{2u^3\ub^3}
\ln^2 u \ln^2 \ub
\no \\
&\quad
+ \frac{2(3 - 116u + 374u^2 - 465u^3 + 280u^4 - 78u^5 + 17u^6 -
  16u^7+4u^8)}{u^3\ub^3} \Li_3(u)
\no \\
&\quad
- \frac{6 - 116u + 149u^2 - 50u^3 - 6u^4 + 6u^5 - 4u^6}{2u^3\ub} \ln^2 u
\ln\ub
\no \\
&\quad
+ \frac{2(3 - 3u - 7u^2 - 3u^3)}{u^3} \bigg( \Li_3(-u) - \ln u \,
\Li_2(-u) - \frac{\ln^2 u+\pi^2}{2}\ln(1+u) \bigg)
+ (\ub-u)
\no \\
&\quad
\times
\bigg(\frac{92 - 117u + 109u^2 + 16u^3 - 8u^4}{4u^2\ub^2}
+ \frac{(1 - 2u+2u^2)(83 - 14u + 14u^2) \pi^2}{6u^3\ub^3}\bigg) \Li_2(u)
\no \\
&\quad
+ \!\!\bigg( \frac{46 - 73u + 26u^2 + 13u^3 - 4u^4}{2u\ub^2} - 4 \pi^2
\!\bigg)\!\ln^2 u 
+ \!\!\bigg( \frac{92 - 289u + 421u^2 - 264u^3 + 132u^4}{4u\ub^2}
\no \\
&\quad
 + \frac{(139 - 554u + 856u^2- 558u^3 + 164u^4 + 138u^5 -
   46u^6)\pi^2}{3u^2\ub^3}\bigg) \ln u \ln \ub
\no \\
&\quad
- \bigg( \frac{183 - 308u}{2\ub} - \frac{(96 - 160u + 55u^2 + 15u^3 +
  3u^4 -2u^5)\pi^2}{3u^2\ub}\bigg)\ln u
\no \\
&\quad
+ \frac{(191 - 784u + 1262u^2 - 1121u^3 + 973u^4 - 495u^5 + 165u^6)
  \pi^4}{180u^3\ub^3}
\no \\
&\quad
- \frac{(580 - 1763u + 2239u^2 - 952u^3 + 476u^4)\pi^2}{48u^2\ub^2}
\no \\
&\quad
- \frac{3(1 - 37u+ 119u^2 - 138u^3 + 4u^4 + 78u^5 - 26u^6)
  \zeta_3}{u^3\ub^3} 
+ \frac{1659}{4}
+ (u\leftrightarrow\ubar)\bigg]
\no \\
&
+ \bigg[
\frac{(1 + u)(1 + 4u - 7u^2)}{3u\ub^2} \bigg( 12\calH_1(u) +
\pi^2\ln(1+u)\bigg)
- 36\ln u \, \Li_3(u)
+ 14\ln^2 u\, \Li_2(u)
\no \\
&\quad
+ \frac{2(1 - 4u + 3u^2 - u^3)}{3\ub^3} \bigg( 24 \calH_2(u)- 2\pi^2
\Li_2(-u)\bigg)
+ \frac{4(2 - 23u + 18u^2 - 6u^3)}{\ub^3} \Li_4(u)
\no \\
&\quad
- \frac{4(52 - 152u + 156u^2 - 61u^3 + 18u^4 - u^5)}{u^3\ub^2}
\S_{2,2}(u)
- \frac{(\ub - u)(13 - 22u \ub)}{u^3\ub^3} \Li_2(u)^2
\no \\
&\quad
- \frac{2(13 - 51u + 77u^2 - 59u^3 + 13u^4 - 7u^5 + u^6)}{u^3\ub^3}
\bigg( 8\ln\ub \, \Li_3(u) + \ln^2 u \ln^2 \ub - 8\zeta_3 \ln u\bigg)
\no \\
&\quad
- \frac{13 - 48u + 66u^2 - 30u^3 + 18u^5 - 6u^6}{u^3\ub^3}  \bigg(\ln u
\ln\ub - \frac{7 \pi^2}{6} \bigg) \Li_2(u)
- \ln^4 u
+ \frac83\ln^3 u \ln \ub
\no \\
&\quad
+ \frac{2(3 - 110u + 339u^2 - 362u^3 + 100u^4 + 59u^5 - 46u^6 +
  14u^7)}{u^3\ub^3} \Li_3(u)
\no \\
&\quad
- \frac{6 - 156u + 227u^2 - 57u^3 + 38u^4}{u^3\ub} \ln u \,\Li_2(u)
- \frac{9 - 32u + 14u^2}{6\ub} \ln^3 u
\no \\
&\quad
- \frac{6 - 104u + 165u^2 - 54u^3 + 34u^4 - 14u^5}{2u^3\ub} \ln^2 u \ln
\ub
\no \\
&\quad
+ \frac{2(3 + 3u + 5u^2 + 3u^3)}{u^3} \bigg( \Li_3(-u) - \ln u \,
\Li_2(-u) - \frac{\ln^2 u+\pi^2}{2}\ln(1+u) \bigg)
\no \\
&\quad
+ \bigg( \frac{92 - 381u + 161u^2 + 440u^3 - 220u^4}{4u^2\ub^2} -
\frac{4(1 - 3u + 3u^2)(\ub + 2u^3 - u^4)\pi^2}{3u^3\ub^3}\bigg)
\Li_2(u)
\no \\
&\quad
+ \bigg( \frac{46 - 47u - 62u^2 + 55u^3}{2u\ub^2} - \frac{4 \pi^2}{3}
\bigg) \ln^2 u
+ \frac{(\ub - u)(13 + 169u^2 - 8u^3 + 4u^4)\pi^4}{90u^3\ub^3}
\no \\
&\quad
+ \bigg( \frac{(1 + u)(246 - 169u^2)}{4u^2\ub^2} + \frac{(87 - 210u^2 -
  16 u^4)\pi^2}{3u^2\ub^3} \bigg)\ln u \ln \ub
\no \\
&\quad
- \bigg(\frac{93}{6\ub} - \frac{(96 - 112u - 31u^2 + 34u^3 -  7u^4)
\pi^2}{3u^2\ub} + 4\zeta_3\bigg)\ln u
\no \\
&\quad
- \frac{(\ub - u)(580 - 187u + 187u^2)\pi^2}{48u^2\ub^2}
+ \frac{8(\ub - u) \pi^2}{u\ub}\ln 2
\no \\
&\quad
- \frac{3(\ub - u)(1 - 39u + 42u^2 - 6u^3 + 3u^4)\zeta_3}{u^3\ub^3}
- (u\leftrightarrow\ubar) \bigg],
\no \\
&h_8(u) =
\bigg[
-\frac{(1 + u)(3 - 4u + 3u^2)}{6u\ub^2} \bigg( 12\calH_1(u) +
\pi^2\ln(1+u)\bigg)
+ 18\ln u\, \Li_3(u)
\no \\
&\quad
- \frac{u}{3\ub^3} \bigg( 24 \calH_2(u)- 2\pi^2\Li_2(-u)\bigg)
- \frac{5 - 22u + 20u^2 - 6u^3}{\ub^3} \Li_4(u)
- 3\ln^2 u \,\Li_2(u)
\no \\
&\quad
+ \frac{4(5 - 11u + 5u^2 + 5u^3 + 4u^4 - u^5)}{u^3\ub^2} \S_{2,2}(u)
- \frac12\ln^4 u
+ \ln^3 u \ln \ub
\no \\
&\quad
+ \frac{4(5 - 16u + 16u^2 - u^3 + 3u^4 - 3u^5 + u^6)}{u^3\ub^3} \bigg(
\ln \ub \, \Li_3(u) - \zeta_3 \ln u\bigg)
\no \\
&\quad
+ \frac{(\ub - u)^2(5 - 3u + 9u^2 - 12u^3 + 6u^4)}{4u^3\ub^3} \Li_2(u)^2
+ \frac{3 + u - 3u^2 + 2u^3}{6\ub} \ln^3 u
\no \\
&\quad
+ \frac{(\ub - u) (5 - 13u + 15u^2 - 4u^3 + 2u^4)}{4u^3\ub^3} \ln u
\,\ln\ub \,\Li_2(u) 
\no \\
&\quad
+ \frac{10 - 32u + 32u^2 + 7u^3 - 21u^4 + 21u^5 - 7u^6}{4u^3\ub^3} \ln^2
u\ln^2 \ub
\no \\
&\quad
- \frac{6 - 61u + 157u^2 - 196u^3 + 134u^4 - 42u^5 + 32u^6 - 32u^7 +
  8u^8}{2u^3\ub^3} \Li_3(u)
\no \\
&\quad
+ \frac{6 - 39u + 47u^2 - 35u^3 - 12u^4}{2u^3\ub} \ln u \,\Li_2(u)
- \frac{(4 + 35u - 254u^2 + 438u^3 - 219u^4)\pi^2}{48u^2\ub^2}
\no \\
&\quad
+ \frac{(2 - u + u^2)(3 - 13u + 5u^2 + 2u^3 - 4u^4)}{4u^3\ub} \ln^2 u
\ln \ub
\no \\
&\quad
- \frac{6 - 3u - 8u^2 - 3u^3}{2u^3} \bigg( \Li_3(-u) - \ln u \,
\Li_2(-u) -\frac{\ln^2 u+\pi^2}{2}\ln(1+u) \bigg)
- (\ub-u)
\no \\
&\quad
\times
 \bigg( \frac{(4 - 7u + 2u^2)(1 - 3u - 2u^2)}{4u^2\ub^2} + \frac{(19 -
  59u + 73u^2 - 28u^3 + 14u^4)\pi^2}{24u^3\ub^3} \bigg)\Li_2(u)
\no \\
&\quad
- \bigg( \frac{8 + 37u - 136u^2 + 115u^3 - 8u^4}{8u\ub^2} - \frac{3
  \pi^2}{2} \bigg) \ln^2 u
\no \\
&\quad
- \bigg( \frac{4 - 3u + 24u^2 - 42u^3 + 21u^4}{8u^2\ub^2}
+ \frac{(127 - 590 u^2 + 652 u^4 - 128 u^6)  \pi^2}{48u^3\ub^3}\bigg)
\ln u \ln \ub
\no \\
&\quad
+ \bigg(\frac{1433 - 2710u}{24\ub} - \frac{(12 - 30u + 8u^2 + 11u^3 +
  3u^4 -2u^5)\pi^2}{6u^2\ub}\bigg) \ln u
\no \\
&\quad
- \frac{(23 - 102u + 176u^2 - 273u^3 + 449u^4 - 375u^5 +
  125u^6)\pi^4}{360u^3\ub^3}
\no \\
&\quad
+ \frac{(6 - 57u + 134u^2 + 26u^3 - 463u^4 + 540u^5 -
  180u^6)\zeta_3}{4u^3\ub^3} 
- \frac{12641}{48}
+ (u\leftrightarrow\ubar)\bigg]
\no \\
&
+ \bigg[
-\frac{(1 + u)(1 + 4u - 7u^2)}{6u \ub^2}\bigg( 12\calH_1(u) +
\pi^2\ln(1+u)\bigg)
+ 2\ln u \,\Li_3(u)
- 3\ln^2 u \,\Li_2(u)
\no \\
&\quad
- \frac{1 - 4u + 3u^2 - u^3}{3\ub^3} \bigg( 24
\calH_2(u)-2\pi^2\Li_2(-u)\bigg) 
+ \frac{19 - 44u + 36u^2 - 14u^3}{\ub^3} \Li_4(u)
\no \\
&\quad
+ \frac{4(5 - 11u + 5u^2 + 5u^3 + 4u^4 - u^5)}{u^3\ub^2} \S_{2,2}(u)
+ \frac{(\ub - u)(\ub + u^2)(5 + 6u - 6u^2)}{4u^3\ub^3}\Li_2(u)^2
\no \\
&\quad
+ \frac{5 - 16u + 16u^2 - 5u^3 - 7u^4 + u^5 + u^6}{2u^3\ub^3}\bigg(
8\ln\ub \, \Li_3(u) + \ln^2 u \ln^2 \ub - 8\zeta_3 \ln u\bigg)
\no \\
&\quad
+ \frac{3 - 6 u^2 + 36 u^4 - 16 u^6}{8u^3\ub^3}\bigg(\ln u \ln\ub -
\frac{7\pi^2}{6} \bigg) \Li_2(u) 
 - \frac{1}{12} \ln^4 u
- \frac23\ln^3 u \ln \ub
\no \\
&\quad
- \frac{(\ub - u)(7 - 4u - 20u^2 + 48u^3 - 24u^4)}{16u^3\ub^3} \ln^2 u
\ln^2 \ub
\no \\
&\quad
- \frac{18 - 165u + 237u^2 + 242u^3 - 702u^4 + 546u^5 - 278u^6 + 84
  u^7}{6u^3\ub^3} \Li_3(u)
\no \\
&\quad
+ \frac{6 - 33u + 19u^2 + 7u^3 + 48u^4}{2u^3\ub} \ln u \,\Li_2(u)
+ \frac{30 - 42u + 7u^2}{6\ub} \ln^3 u
\no \\
&\quad
+ \frac{18 - 69u + 78u^2 - 124u^3 + 217u^4 - 42u^5}{12u^3\ub} \ln^2 u
\ln\ub
\no \\
&\quad
- \frac{6 + 3u + 4u^2 + 3u^3}{2u^3} \bigg( \Li_3(-u) - \ln u \,
\Li_2(-u) -\frac{\ln^2 u+\pi^2}{2}\ln(1+u) \bigg)
\no \\
&\quad
- \frac{(\ub - u)(6 - 283u + 283u^2)\pi^2}{72u^2\ub^2}
- \bigg( \frac{36 - 312u - 1493u^2 + 3610u^3-1805u^4}{36u^2\ub^2}
\no \\
&\quad
- \frac{(2 - 8u + 12u^2 - 5u^3-5u^4+ 9u^5 -
  3u^6)\pi^2}{3u^3\ub^3}\bigg) \Li_2(u)
- \frac{4(\ub - u) \pi^2}{u\ub}\ln2
\no \\
&\quad
- \bigg( \frac{72 + 1703u - 3724u^2 + 1805u^3}{72u\ub^2} +
\frac{\pi^2}{3}\bigg) \ln^2 u
 \no \\
&\quad
- \bigg( \frac{(\ub - u)(3 - 10u + 10u^2)}{6u^2\ub^2} - \frac{(33 - 21u
  + 35u^2 - 28u^3 + 14u^4)\pi^2}{12u^2\ub^3} \bigg)\ln u \ln \ub
 \no \\
&\quad
+ \bigg(\frac{1009}{72\ub} - \frac{(12 + 26u - 121u^2 + 80u^3 -
  7u^4)\pi^2}{6u^2\ub} + 4\zeta_3\bigg)\ln u
  \no \\
&\quad
- \frac{(\ub - u)(429 + 84u - 1244u^2 + 2320u^3 -
  1160u^4)\pi^4}{2880u^3\ub^3}
  \no \\
&\quad
+ \frac{3(\ub - u)(2 - 21u - 4u^2 + 50u^3 - 25u^4) \zeta_3}{4u^3\ub^3}
- (u\leftrightarrow\ubar) \bigg].
\end{align}
The definition of the functions $\calH_{1,2}(x)$ can be found in
Section~\ref{sec:2loop}. The diagrams with a closed fermion loop give
for massless internal quarks 
\begin{align}
&h_9(u;0) =
\bigg[ \frac{125}{12} + \frac{\Li_2(u) }{\ubar} + \frac{1-3u}{2\ubar}
\ln^2 u + \frac{1+u}{2\ubar} \ln u \ln \ubar -\frac{17(1-2u)}{6\ubar}
\ln u
\no \\
&\quad
- \frac{(1+u)\pi^2}{12\ubar} + (u\leftrightarrow\ubar)\bigg]
\no \\
&
+ \bigg[ \frac43 \Li_3(u) -\frac23 \ln^3 u + \frac43 \ln^2 u \ln \ubar -
\frac{32-29u}{9\ubar} \Li_2(u)  + \frac{35-29u}{18\ubar} \ln^2 u
\no \\
&\hspace{12mm}
- \frac{1}{3\ubar} \ln u \ln \ubar - \frac{13+24\ubar \pi^2}{18\ubar}
\ln u + \frac{\pi^2}{18\ubar} - (u\leftrightarrow\ubar) \bigg],
\end{align}
and for an internal $b$-quark
\begin{align}
&h_9(u;1) =
\bigg[
\frac{8}{u^2} \bigg( \Li_3(-x_b) - \S_{1,2}(-x_b) - \ln(1+x_b)
\Li_2(-x_b) - \frac{1}{12} \ln^3 \frac{x_b}{1+x_b}
\no \\
&\quad
- \frac16 \ln^3 (1 + x_b)
+ \frac{\pi^2}{6} \ln \frac{x_b}{1+x_b}  \bigg)
- \frac{14 - 75u^2 + 60u^4 + 19u^6}{9u^3\ub^3} \ln u \ln \ub
\no \\
&\quad
- \frac{2y_b(6 + u)}{u} \bigg( \Li_2(-x_b) - \frac14 \ln^2
\frac{x_b}{1+x_b} + \frac12 \ln^2 (1 + x_b) + \frac{\pi^2}{6} \bigg)
+ \frac{4 u}{\ub^3} \, \Li_3(u)
\no \\
&\quad
- \frac{32 - 204u + 504u^2 - 584u^3 + 405u^4 - 150u^5 +29u^6}{9u^3\ub^3}
\Li_2(u)
- \frac{17(\ub - u)}{6\ub} \ln u
\no \\
&\quad
- \frac{(40 - 213u^2 + 120u^4 - 19u^6)\pi^2}{54u^3\ub^3}
- \frac{2(1 - 6 u^2 + 6u^4)\zeta_3}{u^3\ub^3}
\no \\
&\quad
+ \frac{5(61 - 50u^2)}{24u\ub}
+ (u\leftrightarrow\ubar)\bigg]
\no \\
&
+ \bigg[
-\frac{8(3 - u^2)}{3u^2} \bigg( \Li_3(-x_b) - \S_{1,2}(-x_b) - \ln(1+x_b)
\Li_2(-x_b) - \frac{1}{12} \ln^3 \frac{x_b}{1+x_b}
\no \\
&\quad
- \frac16 \ln^3 (1 + x_b)
+ \frac{\pi^2}{6} \ln \frac{x_b}{1+x_b}  \bigg)
+ \frac{54 - 103 u^2 + 81 u^4}{9u^2\ub^3} \ln u \ln \ub
\no \\
&\quad
+ \frac{2y_b(38 + 29u)}{9u} \bigg( \Li_2(-x_b) - \frac14 \ln^2
\frac{x_b}{1+x_b} + \frac12 \ln^2 (1 + x_b) + \frac{\pi^2}{6} \bigg)
\no \\
&\quad
- \frac{4(1 + 3u^2 - u^3)}{3\ub^3} \Li_3(u)
- \frac{128 - 504u + 389u^2}{18u^2\ub} \ln u
\no \\
&\quad
- \frac{32 - 204u + 504u^2 - 584u^3 + 405u^4 - 150u^5 +29u^6}{9u^3\ub^3}
\Li_2(u)
\no \\
&\quad
+ \frac{(42 - 49 u + 39 u^3)\pi^2}{54u\ub^3}
+ \frac{4\zeta_3}{u^2\ub^3}
+ \frac{261 - 325u^2}{9u\ub^2}
- (u\leftrightarrow\ubar) \bigg],
\end{align}
where we introduced the shorthand notation
\begin{align}
x_b =\frac12 (y_b - 1),\qquad
y_b = \sqrt{\frac{4+u}{u}}.
\end{align}
We finally refrain from presenting the charm quark contribution, which
is rather complicated and depends on two parameterizations for the
4-topology Master Integrals that we could not solve in a closed
analytical form (cf.~the discussion in~\cite{GB:Inclusive}). We may
still evaluate these Master Integrals numerically in
Section~\ref{sec:convolutions} to perform the convolution with the
light-cone distribution amplitude of the emitted meson $M_2$.

\end{appendix}



\begin{thebibliography}{99}

\bibitem{BBNS}
  M.~Beneke, G.~Buchalla, M.~Neubert and C.~T.~Sachrajda,
  Phys.\ Rev.\ Lett.\  {\bf 83} (1999) 1914
  [arXiv:hep-ph/9905312];\\
  M.~Beneke, G.~Buchalla, M.~Neubert and C.~T.~Sachrajda,
  Nucl.\ Phys.\  B {\bf 591} (2000) 313
  [arXiv:hep-ph/0006124];\\
  M.~Beneke, G.~Buchalla, M.~Neubert and C.~T.~Sachrajda,
  Nucl.\ Phys.\  B {\bf 606} (2001) 245
  [arXiv:hep-ph/0104110].


\bibitem{SCET}
  C.~W.~Bauer, S.~Fleming, D.~Pirjol and I.~W.~Stewart,
  Phys.\ Rev.\  D {\bf 63} (2001) 114020
  [arXiv:hep-ph/0011336];\\
  C.~W.~Bauer, D.~Pirjol and I.~W.~Stewart,
  Phys.\ Rev.\  D {\bf 65} (2002) 054022
  [arXiv:hep-ph/0109045];\\
  M.~Beneke, A.~P.~Chapovsky, M.~Diehl and T.~Feldmann,
  Nucl.\ Phys.\  B {\bf 643} (2002) 431
  [arXiv:hep-ph/0206152].


\bibitem{SpecScat:NLO:tree}
  M.~Beneke and S.~Jager,
  Nucl.\ Phys.\  B {\bf 751} (2006) 160
  [arXiv:hep-ph/0512351];\\
  N.~Kivel,
  JHEP {\bf 0705} (2007) 019
  [arXiv:hep-ph/0608291];\\
  V.~Pilipp,
  PhD thesis, LMU M\"unchen, 2007,
  arXiv:0709.0497 [hep-ph];\\
  V.~Pilipp,
  Nucl.\ Phys.\  B {\bf 794} (2008) 154
  [arXiv:0709.3214 [hep-ph]].


\bibitem{SpecScat:NLO:penguin}
  M.~Beneke and S.~Jager,
  Nucl.\ Phys.\  B {\bf 768} (2007) 51
  [arXiv:hep-ph/0610322];\\
  A.~Jain, I.~Z.~Rothstein and I.~W.~Stewart,
  arXiv:0706.3399 [hep-ph].


\bibitem{GB:Im}
  G.~Bell,
  Nucl.\ Phys.\  B {\bf 795} (2008) 1
  [arXiv:0705.3127 [hep-ph]].


\bibitem{GB:thesis}
  G.~Bell,
  PhD thesis, LMU M\"unchen, 2006,
  arXiv:0705.3133 [hep-ph].


\bibitem{Chetyrkin:1997gb}
  K.~G.~Chetyrkin, M.~Misiak and M.~Munz,
  Nucl.\ Phys.\  B {\bf 520} (1998) 279
  [arXiv:hep-ph/9711280].


\bibitem{Gorbahn:2004my}
  M.~Gorbahn and U.~Haisch,
  Nucl.\ Phys.\  B {\bf 713} (2005) 291
  [arXiv:hep-ph/0411071].


\bibitem{Buras:2006gb}
  A.~J.~Buras, M.~Gorbahn, U.~Haisch and U.~Nierste,
  JHEP {\bf 0611} (2006) 002
  [arXiv:hep-ph/0603079].


\bibitem{Martin}
  M.~Gorbahn, private communication.


\bibitem{IBP}
  F.~V.~Tkachov,
  Phys.\ Lett.\ B {\bf 100} (1981) 65;\\
  K.~G.~Chetyrkin and F.~V.~Tkachov,
  Nucl.\ Phys.\ B {\bf 192} (1981) 159.


\bibitem{Bonciani:2008wf}
  R.~Bonciani and A.~Ferroglia,
  JHEP {\bf 0811} (2008) 065
  [arXiv:0809.4687 [hep-ph]];\\
  H.~M.~Asatrian, C.~Greub and B.~D.~Pecjak,
  Phys.\ Rev.\  D {\bf 78} (2008) 114028
  [arXiv:0810.0987 [hep-ph]].


\bibitem{Beneke:2008ei}
  M.~Beneke, T.~Huber and X.~Q.~Li,
  Nucl.\ Phys.\  B {\bf 811} (2009) 77
  [arXiv:0810.1230 [hep-ph]].


\bibitem{HPLs}
  E.~Remiddi and J.~A.~M.~Vermaseren,
  Int.\ J.\ Mod.\ Phys.\  A {\bf 15} (2000) 725
  [arXiv:hep-ph/9905237].


\bibitem{GB:Inclusive}
  G.~Bell,
  Nucl.\ Phys.\  B {\bf 812} (2009) 264
  [arXiv:0810.5695 [hep-ph]].


\bibitem{Huber:2009se}
  T.~Huber,
  arXiv:0901.2133 [hep-ph].


\bibitem{Gambino:2003zm}
  P.~Gambino, M.~Gorbahn and U.~Haisch,
  Nucl.\ Phys.\  B {\bf 673} (2003) 238
  [arXiv:hep-ph/0306079].


\bibitem{ERBL}
  A.~V.~Efremov and A.~V.~Radyushkin,
  Phys.\ Lett.\  B {\bf 94} (1980) 245;\\
  G.~P.~Lepage and S.~J.~Brodsky,
  Phys.\ Rev.\  D {\bf 22} (1980) 2157.


\bibitem{betazero}
  T.~Becher, M.~Neubert and B.~D.~Pecjak,
  Nucl.\ Phys.\  B {\bf 619} (2001) 538
  [arXiv:hep-ph/0102219];\\
  C.~N.~Burrell and A.~R.~Williamson,
  Phys.\ Rev.\  D {\bf 73} (2006) 114004
  [arXiv:hep-ph/0504024].


\bibitem{Hill:2004if}
  R.~J.~Hill, T.~Becher, S.~J.~Lee and M.~Neubert,
  JHEP {\bf 0407} (2004) 081
  [arXiv:hep-ph/0404217];\\
  M.~Beneke and D.~Yang,
  Nucl.\ Phys.\  B {\bf 736} (2006) 34
  [arXiv:hep-ph/0508250].


\bibitem{Lee:2005gza}
  S.~J.~Lee and M.~Neubert,
  Phys.\ Rev.\  D {\bf 72} (2005) 094028
  [arXiv:hep-ph/0509350].


\bibitem{Becher:2004kk}
  T.~Becher and R.~J.~Hill,
  JHEP {\bf 0410} (2004) 055
  [arXiv:hep-ph/0408344];\\
  G.~G.~Kirilin,
  arXiv:hep-ph/0508235.


\bibitem{Bobeth:1999mk}
  C.~Bobeth, M.~Misiak and J.~Urban,
  Nucl.\ Phys.\  B {\bf 574} (2000) 291
  [arXiv:hep-ph/9910220].


\bibitem{Tarasov:1980au}
  O.~V.~Tarasov, A.~A.~Vladimirov and A.~Y.~Zharkov,
  Phys.\ Lett.\  B {\bf 93} (1980) 429;\\
  S.~A.~Larin and J.~A.~M.~Vermaseren,
  Phys.\ Lett.\  B {\bf 303} (1993) 334
  [arXiv:hep-ph/9302208].


\bibitem{IRrearrange}
  M.~Misiak and M.~Munz,
  Phys.\ Lett.\  B {\bf 344} (1995) 308
  [arXiv:hep-ph/9409454];\\
  K.~G.~Chetyrkin, M.~Misiak and M.~Munz,
  Nucl.\ Phys.\  B {\bf 518} (1998) 473
  [arXiv:hep-ph/9711266].


\bibitem{Egorian:1978zx}
  E.~Egorian and O.~V.~Tarasov,
  Teor.\ Mat.\ Fiz.\  {\bf 41} (1979) 26
  [Theor.\ Math.\ Phys.\  {\bf 41} (1979) 863].


\bibitem{Tarrach:1980up}
  R.~Tarrach,
  Nucl.\ Phys.\  B {\bf 183} (1981) 384;\\
  D.~J.~Broadhurst and A.~G.~Grozin,
  Phys.\ Rev.\  D {\bf 52} (1995) 4082
  [arXiv:hep-ph/9410240].


\end{thebibliography}
\end{document}